\newcommand{\version}{July 30, 2016}
\documentclass[letterpaper,11pt,oneside,pdftex]{article}
\pdfoutput=1

\usepackage[bindingoffset=0.0cm,textheight=22.6cm,hdivide={*,17.00cm,*}, vdivide={*,22.5cm,*}]{geometry}
\usepackage{latexsym}
\usepackage{amsmath,amssymb}
\usepackage{mathtools}
\IfFileExists{dsfont.sty}
	{\usepackage{dsfont}
         \let\mathbb=\mathds
         \newcommand{\id}{\mathds{1}}}
	{\typeout{Package dsfont.sty was not found, using alternative macros.}
         \let\mathds=\mathbb
         \newcommand{\id}{\mbox{1 \kern-.59em \textrm{l}}}}
\usepackage[T1]{fontenc}
\usepackage[utf8]{inputenc}
\usepackage{lmodern}
\usepackage{slashed}
\usepackage[pdftex]{graphicx}
\IfFileExists{biblatex.sty}
 {%
 \usepackage[backend=bibtex,bibencoding=ascii,sorting=none,sortcites=true,citestyle=numeric-comp,bibstyle=custom,maxnames=6,minnames=3,firstinits=true]{biblatex}
 \addbibresource{neutrinos}%
 }
 {%
 \typeout{Package biblatex.sty was not found, falling back to natbib.}
 \usepackage[numbers,square,comma,sort&compress]{natbib}
 \newcommand{\printbibliography}{\bibliographystyle{utphys-custom} \bibliography{neutrinos}}%
 }
\usepackage[pdftex,hyperref,svgnames]{xcolor}
\usepackage[pdftex,bookmarksnumbered=true,breaklinks=true]{hyperref}
\makeatletter
\AtBeginDocument{
  \hypersetup{
    pdftitle = {\@title},
    pdfsubject = {\Abstract},
    pdfauthor = {\@author}
  }
}
\makeatother
\usepackage{tensind}
\tensordelimiter{?}
\makeatletter

 \@addtoreset{equation}{section}
 \makeatother

\renewcommand{\a}{\alpha}
\renewcommand{\b}{\beta}
\newcommand{\g}{\gamma}
\renewcommand{\d}{\delta}
\newcommand{\e}{\epsilon}

\renewcommand{\th}{\theta}

\renewcommand{\l}{\lambda}
\newcommand{\m}{\mu}
\newcommand{\n}{\nu}

\renewcommand{\r}{\rho}
\newcommand{\s}{\sigma}

\DeclarePairedDelimiter\abs{\lvert}{\rvert}

\newcommand{\eqnref}[1]{Eqn.~(\ref{#1})}

\newcommand{\appref}[1]{Appendix~\ref{#1}}

\newcommand{\Tr}{\textrm{Tr}}
\newcommand{\tr}{\textrm{tr}}

\newcommand{\beq}{\begin{equation}}
\newcommand{\eeq}{\end{equation}}
\newcommand{\be}{\begin{equation}}
\newcommand{\ee}{\end{equation}}
\newcommand{\bea}{\begin{eqnarray}}
\newcommand{\eea}{\end{eqnarray}}
\newcommand{\nn}{\nonumber}

\newcommand{\sgn}{\textrm{sgn}}
\def\greaterthansquiggle{\raise.3ex\hbox{$>$\kern-.75em\lower1ex\hbox{$\sim$}}}
\def\lessthansquiggle{\raise.3ex\hbox{$<$\kern-.75em\lower1ex\hbox{$\sim$}}}

\newcommand{\ba}{\begin{array}}
\newcommand{\ea}{\end{array}}

\newcommand{\vp}{\varphi}

\newcommand{\cL}{{\cal L}}
\newcommand{\M}{{\cal M}}

\def\au{{\setbox0=\hbox{\lower1.36775ex%
\hbox{''}\kern-.05em}\dp0=.36775ex\hskip0pt\box0}}
\def\ao{{}\kern-.10em\hbox{``}}

\title{\texorpdfstring{\begin{flushright}
        {\small LA-UR-15-25029}
       \end{flushright}\vspace{2em}}{}%
       \textbf{Neutrino Quantum Kinetic Equations:\texorpdfstring{\\}{} The Collision Term}}
\author{Daniel N. Blaschke and  Vincenzo Cirigliano}
\date{\version}
\newcommand{\Abstract}{%
We derive the collision term relevant for neutrino quantum kinetic equations in the early universe and compact astrophysical objects,
displaying its full matrix structure in both flavor and spin degrees of freedom.  We include in our analysis neutrino-neutrino processes,
scattering and annihilation with electrons  and positrons, and neutrino scattering off nucleons (the latter in the low-density limit).
After presenting the general structure of the collision terms,  we take two instructive limiting cases.
The one-flavor limit  highlights the structure in helicity space and  allows for  a straightforward interpretation 
of the off-diagonal entries in terms of the product of scattering amplitudes of the two helicity states.
The isotropic limit is relevant for studies of the early universe: in this case the terms involving
spin coherence vanish and the collision term can be expressed
in  terms of two-dimensional integrals, suitable for computational implementation.
}

\graphicspath{{./}{./figures/}}
\begin{document}
\maketitle
\thispagestyle{empty}
\begin{center}
\vspace{-0.3cm}
Los Alamos National Laboratory\\Los Alamos, NM, 87545, USA
\\[0.5cm]
\ttfamily{E-mail: dblaschke@lanl.gov, cirigliano@lanl.gov}
\end{center}

\vspace{1.5em}
\begin{abstract}
\Abstract
\end{abstract}

\newpage
\setcounter{tocdepth}{2}
\tableofcontents

\section{Introduction}
\label{sect:intro}

The evolution of  an ensemble of neutrinos in hot and dense media 
is described by an appropriate set of quantum kinetic equations (QKEs), 
accounting for kinetic, flavor, and  spin degrees of 
freedom~\cite{Sigl:1992fn,Raffelt:1992uj,McKellar:1994uq,Barbieri:1991fj,Enqvist:1990ad,Rudzsky:1990,Strack:2005fk,Volpe:2013lr,Vlasenko:2013fja,Zhang:2013lka,
Cirigliano:2014aoa,Serreau:2014cfa,Kartavtsev:2015eva,Dobrynina:2016rwy}.   
QKEs are central to obtain a complete 
description of  neutrino transport in the early universe, 
core collapse supernovae, and compact object mergers, 
valid before, during, and after the neutrino decoupling epoch (region).  
A self-consistent treatment of neutrino transport is highly relevant because in  
such environments  neutrinos carry a significant fraction of the energy and entropy, and through their flavor- and energy-dependent 
weak interactions play a key role in setting the neutron-to-proton ratio, a critical input for the nucleosynthesis process. 

In Ref.~\cite{Vlasenko:2013fja}  the  QKEs describing the evolution of Majorana neutrinos  
were derived using field-theoretic methods (see~\cite{Schwinger:1960qe,Keldysh:1964ud,Craig:1968,Calzetta:1988qy,Berges:2004yj} for an introduction to non-equilibrium QFT). These QKEs include spin degrees of freedom and encompass effects up to second order in small ratios of scales  
characterizing the neutrino environments we are interested in. 
Specifically,  we treat  neutrino masses,  mass-splitting,  and matter potentials induced by forward scattering,  
as well as  external gradients  as much smaller than 
the typical neutrino energy scale $E$, set by the temperature or chemical potential: namely  $m_\nu/E \sim \Delta m_\nu/E \sim  \Sigma_{\rm forward}/E 
\sim \partial_X/E \sim O(\epsilon)$~\footnote{In the early universe, the small lepton number implies  $\Sigma_{\rm forward}  \sim G_F n_e  \ll m_\nu \sim \Delta m_\nu$. 
This is not the case in supernovae.}.  
The inelastic  scattering can also be characterized by a potential  $\Sigma_{\rm inelastic} \sim \Sigma_{\rm forward} \times G_F  E^2$
which we therefore   power-count as  $\Sigma_{\rm inelastic}/E \sim O(\epsilon^2)$. 
This power-counting  is tantamount to the  statement  that physical quantities 
vary slowly on the scale of the neutrino de Broglie wavelength. 

In this paper we elaborate on the terms of the QKEs describing  inelastic collisions or production and absorption in the medium. 
These terms are essential for a correct description of neutrinos in the decoupling epoch (region),  in which the neutrino spectra and 
flavor composition are determined~\cite{Hannestad:1995rs,Dolgov:1997mb,Mangano:2005cc,Grohs:2015tfy,Grohs:2015eua}.
While Ref.~\cite{Vlasenko:2013fja} only included a discussion of neutrino-neutrino scattering in isotropic environment, 
here  we compute the collision terms induced by 
neutrino-(anti)neutrino processes, neutrino scattering and annihilation with electrons  and positrons,  
and neutrino scattering off nucleons.  
Our expressions for  processes involving nucleons are valid in the low-density limit, i.e.  do not 
take into account nucleon interactions.
However, the  effects of strong interactions in dense matter  ---
relevant for supernovae environments --- can be included by appropriately modifying the
medium response functions  (see for example \cite{Reddy:1997yr,Bruenn:1985en,Tubbs:1975jx}).

The paper is organized as follows: in Sect.~\ref{sect:review} we review the Quantum Kinetic Equations (QKEs)
in the field-theoretic approach,  
for both Dirac and Majorana neutrinos.
In Sect.~\ref{sect:derivation} we provide a derivation of the generalized collision term 
for Majorana neutrinos, and present  general expressions involving coherence terms both in flavor and spin space, 
valid for any geometry, relegating some lengthy results to Appendix~\ref{app:results}.
The collision terms for Dirac neutrinos  are discussed in Section~\ref{app:dirac}.
After presenting general results, we discuss two limiting  cases. 
In Sect.~\ref{sect:oneflavor} we take the one-flavor limit and illustrate the structure of the collision term for the two spin degrees of freedom 
of a  Majorana neutrino (i.e neutrino and antineutrino). 
In Sect.~\ref{sect:isotropic} we consider the isotropic limit relevant for the description of neutrinos in the early universe, 
with some details reported in Appendix~\ref{sec:DHSintegrals}.
Finally, we present our concluding remarks in Sect.~\ref{sect:conclusion}.
To keep the paper self-contained, we include a number of appendices with technical details and lengthy results. 

\section{Review of quantum kinetic equations (QKEs)}
\label{sect:review}

In this section we review the field-theoretic approach to neutrino QKEs, following Refs.~\cite{Vlasenko:2013fja,Cirigliano:2014aoa}, 
with the dual purpose of having a self-contained presentation and setting the notation for the following sections. 
After a brief discussion of neutrino interactions in the Standard Model (SM) at energy scales much smaller than the $W$ and $Z$ boson masses, 
we   present  the Green's function approach to neutrino propagation in hot and dense media, 
we  describe the structure of the QKEs,  and  we finally  review the content of the ``coherent'' (collisionless) QKEs. 
Throughout, we use  four-component spinors to describe both Dirac and  Majorana neutrinos, providing 
an alternative description to the one of Ref.~\cite{Vlasenko:2013fja}, that employs two-dimensional Weyl spinors.
In this section we present results for both Majorana and Dirac neutrinos.

\subsection{Neutrino interactions}
\label{sect:int}

In this work we describe neutrino fields (Dirac or Majorana) in terms of 
4-component spinors $\nu_\alpha$, where  $\alpha$ is a flavor or family index.   
In the Majorana case the fields satisfy the 
Majorana condition  $\nu^c = \nu$,  with $\nu^c \equiv  C \bar{\nu}^T$, 
where  $C = i  \gamma_0 \gamma_2$ is the charge-conjugation matrix. 
In the Majorana case, the kinetic Lagrangian can be written as 
\be
{\cal L}_{\rm Kin} =  \frac{i}{2} \,  \bar{\nu} \slashed{\partial}  \nu  \ - \ \frac{1}{2}  \bar{\nu}  \, m \, \nu ~, 
\qquad \qquad 
\nu = 
\left(  \begin{array}{c} 
\nu_e \\
\nu_\mu\\
\nu_\tau
\end{array}
\right)~, 
 \qquad  
\ee
where $m = m^T$ is the Majorana mass matrix (a complex symmetric matrix).~\footnote{
For Dirac neutrinos the kinetic Lagrangian reads
${\cal L}_{\rm Kin} =   i \,  \bar{\nu} \slashed{\partial}  \nu  \ - \  \bar{\nu}  \, m \, \nu$, 
where $m$ is a generic complex matrix.}  

In situations of  physical interest, such as neutrino decoupling in the early universe 
and neutrino propagation in compact astrophysical objects, the typical
neutrino energy is well below the electroweak scale ($\sim 100$~GeV). 
Therefore, in computing the collision integrals it is safe to use the 
contact-interaction limit of the full Standard Model, and to replace the 
quark degrees of freedom with nucleon degrees of freedom.  

After integrating out $W$ and $Z$ bosons,   the part of the Standard Model 
effective Lagrangian controlling neutrino interactions can be written in the following current-current form  
(in terms of 4-dimensional spinors): 
\begin{subequations}\label{eq:interaction-lagrangian}
\begin{align}
{\cal L}_{\nu \nu} &=  - \frac{G_F}{\sqrt{2}} \  \bar{\nu}  \gamma_\mu P_L  \nu \  \,  \bar{\nu}  \gamma^\mu P_L  \nu \,, \\
{\cal L}_{\nu e} &=  - 2 \sqrt{2}   G_F\  
\Big(  \bar{\nu}  \gamma_\mu P_L  Y_{L}   \nu \  \,  \bar{e}  \gamma^\mu P_L e  
\  +  \   \bar{\nu}  \gamma_\mu P_L  Y_{R}   \nu \  \,  \bar{e}  \gamma^\mu P_R e   \Big) 
\,,\\
{\cal L}_{\nu N} &=  - \sqrt{2} G_F \   \sum_{N=p,n}  \   \bar{\nu}  \gamma_\mu P_L   \nu \  \, 
\bar{N} \gamma^\mu  \left(C_V^{(N)}  -  C_A^{(N)} \gamma_5\right) N  
\,,\\
{\cal L}_{CC} &=  - \sqrt{2} G_F \   \bar{e}  \gamma_\mu P_L \nu_e \   \bar{p}\,   \gamma^\mu \left(1 - g_A \gamma_5\right)  n   \ + \ {\rm h.c.} \,,
\end{align}
\end{subequations}
where $G_F$ is the Fermi constant, $P_{L,R} =  (\id \mp  \gamma_5)/2$, and 
\be
Y_{L}  = \left(  \begin{array}{ccc} 
\frac{1}{2}  + \sin^2 \theta_W   &0&0 \\
0 & - \frac{1}{2}  + \sin^2 \theta_W  &0\\
0 & 0 & 
- \frac{1}{2}  + \sin^2 \theta_W 
\end{array}
\right)~, 
\quad 
Y_{R}  =   \sin^2 \theta_W \times \id~. 
\ee
The nucleon couplings are given in terms of   $g_A \simeq 1.27$ by 
\begin{align}
C_V^{(p)} &= \frac{1}{2} - 2 \sin^2 \theta_W    \,,   &
C_V^{(n)} &=-  \frac{1}{2}  \,,  \nn\\
C_A^{(p)} &= \frac{g_A}{2}   \,, &
C_A^{(n)} &=-  \frac{g_A}{2} \,.
\label{eq:nucleoncouplings}
\end{align}

\subsection{Neutrinos in  hot and dense media}
QKEs are the evolution equations for suitably defined dynamical quantities 
that characterize a neutrino ensemble, which we will refer to as neutrino density matrices. 
In  the most general  terms a neutrino ensemble is described by 
the set  of all $2n$-field Green's functions, 
encoding  $n$-particle correlations. 
These obey  coupled  integro-differential equations, 
equivalent to the BBGKY equations~\cite{Calzetta:1988qy}.  
As discussed in Refs.~\cite{Sigl:1992fn,Vlasenko:2013fja}, for weakly interacting neutrinos ($\Sigma/E \sim O(\epsilon, \epsilon^2)$)  
the set of coupled equations  can be truncated by  
using perturbation theory to express
all higher order Green's functions 
in terms of the two-point functions.   
In this case the neutrino ensemble is characterized by one-particle correlations.\footnote{We neglect here 
correlations that pair particles and antiparticles of opposite momenta~\cite{Volpe:2013lr,Serreau:2014cfa}.
The coupling  of  these new densities  to the standard density matrices 
has been worked out explicitly in Ref.~\cite{Serreau:2014cfa}.
We neglect these terms  as their effect primarily  generates coherence  of opposite-momentum neutrinos 
only for  very long-wavelength  modes,  with $\lambda_{\rm de Broglie} \sim \lambda_{\rm scale-hight}$, where  
$\lambda_{\rm scale-hight}$ is the length scale characterizing a given astrophysical environment.      
Significant feedback effects  from the long-wavelength modes could  alter the  analysis presented here.
However,  a detailed study of this point goes beyond the scope of this work.}

One-particle states of  massive neutrinos and antineutrinos are
specified by the three-momentum $\vec{p}$,  the helicity $h \in \{L,R \}$, and  the  family label $i$ (for eigenstates of mass $m_i$),  
with corresponding  annihilation operators  
$a_{i, \vec{p}, h}$ and  $b_{j, \vec{p}, h}$ 
satisfying the canonical anti-commutation relations 
$ \{ a_{i, \vec{p}, h},
a^{\dagger}_{j, \vec{p}', h'} 
\} = (2 \pi)^3  \, 2 \, \omega_i (\vec{p})  \, \delta_{h h'} \, \delta _{ij} \, \delta^{(3)} (\vec{p} - \vec{p}')$, {\it etc.},  
where $\omega_i (\vec{p}) = \sqrt{\vec{p}^2 + m_i^2}$.
Then, the neutrino state is specified by  the matrices  $f_{h h'}^{ij} (\vec{p})$ and 
$\bar{f}_{h h'}^{ij} (\vec{p})$    (which we call density matrices,  with slight abuse of language) 
\begin{subequations}
\label{eq:f1}
\begin{align}
\langle 
a^\dagger_{j, \vec{p}', h'}  \,  a_{i, \vec{p}, h}
 \rangle   &=  (2 \pi)^3 \, 2 n (\vec{p})  \, \delta^{(3)} (\vec{p} - \vec{p}' ) \  f_{h h'}^{ij} (\vec{p}) ~, \quad 
\\
\langle
b^\dagger_{i, \vec{p}', h'}  \,  b_{j, \vec{p}, h}
\rangle   &=  (2 \pi)^3 \, 2 n (\vec{p})  \, \delta^{(3)} (\vec{p} - \vec{p}' ) \  \bar{f}_{h h'}^{ij} (\vec{p}) ~, 
\end{align}
\end{subequations}
where $\langle \dots \rangle$ denotes the ensemble average and  $n (\vec{p})$ is a normalization factor.~\footnote{The interchange $i \leftrightarrow j$ in 
the definition of antiparticle distribution matrices is chosen so that  under unitary transformations   $\nu'  = U \nu$,  $f$ and $\bar{f}$ transform 
in the same way, i.e.  $f ' =      U f  U^\dagger$. } 
For  inhomogeneous backgrounds, the density matrices depend  also on the space-time label, denoted by $x$ in what follows. 

The physical meaning of the generalized 
density matrices $f_{h h'}^{ij} (\vec{p})$ and 
$ \bar{f}_{h h'}^{ij} (\vec{p})$
is  dictated by simple quantum mechanical considerations: 
the diagonal entries   $f_{hh}^{ii} (\vec{p})$ represent the occupation numbers 
 of  neutrinos of mass $m_i$, momentum $\vec{p}$, and helicity $h$;  
the off diagonal elements    $f_{hh}^{ij} (\vec{p})$  represent  quantum coherence  of states of same helicity and  different mass 
(familiar in the context of neutrino oscillations);  $f_{hh'}^{ii} (\vec{p})$  represent  coherence of  states of different helicity and same mass,  
 and finally   $f_{hh'}^{ij} (\vec{p})$  represent  coherence between states of different helicity and mass.  

In summary,  the basic dynamical objects describing  ensembles of neutrinos  and antineutrinos 
are the $2 n_f \times 2 n_f$ 
matrices, 
\be
F 
(\vec{p}, x) =
\left(
\begin{array}{cc}
f_{LL} & f_{LR}\\
f_{RL} &  f_{RR}
\end{array}
\right);  
\qquad \qquad \bar{F}
(\vec{p}, x) 
= 
\left(
\begin{array}{cc}
\bar{f}_{RR} & \bar{f}_{RL}\\
\bar{f}_{LR} &  \bar{f}_{LL}
\end{array}
\right), 
\label{eq:FFbar1}
\ee
where we have suppressed the generation indices (each block $f_{h h'}$ is a square $n_f \times n_f$ matrix).  
QKEs are the evolution equations for $F$ and $\bar{F}$.  Before sketching their derivation in the following subsections,  we 
discuss how  this formalism allows one to describe both Dirac and Majorana neutrinos:
\begin{itemize}
\item For  Dirac neutrinos, one needs  both $F$ and $\bar{F}$, with $f_{LL}$ and $\bar{f}_{RR}$  denoting the occupation numbers of active states, 
left-handed neutrinos and right-handed antineutrinos, respectively. Similarly,   $f_{RR}$ and $\bar{f}_{LL}$ describe the occupation number of 
wrong-helicity sterile states. 
\item For  Majorana neutrinos,  one can choose the phases so that $a_i (\vec{p},h) = b_i (\vec{p}, h)$  and therefore  
$f_{hh'} = \bar{f}_{hh'}^T$ (transposition acts  on flavor indices). Therefore the ensemble  is  described by just the matrix $F(\vec{p},x)$.  
With the  definitions $f \equiv f_{LL}$,  $ \bar{f} \equiv \bar{f}_{RR} = f_{RR}^T$,  
and $\phi \equiv  f_{LR}$,   one needs evolution equations only for the matrix ${\cal F}$ introduced in Ref.~\cite{Vlasenko:2013fja}:
\be
F \to {\cal F} = 
\left(
\begin{array}{cc}
f &  \phi \\
\phi^\dagger  &  \bar{f}^T
\end{array}
\right)~.
\label{eq:FMajorana}
\ee 
Here $f$ and $\bar{f}$ are $n_f \times n_f$ matrices describing the occupation and flavor coherence of neutrinos and antineutrinos, respectively.
The  $n_f \times n_f$   ``spin coherence'' matrix $\phi$  describes the degree to which the ensemble contains coherent superpositions of neutrinos and antineutrinos of any flavor. 
\end{itemize}

The above discussion in terms of creation and annihilation operators  
has been presented in  the mass eigenstate basis~\cite{Giunti:1991cb}.  
One can define ``flavor basis'' density matrices
$f_{\alpha \beta}$ in terms of the mass-basis $f_{ij}$  as 
$f_{\alpha \beta} = U_{\alpha i} f_{ij} U^*_{\beta j}$, 
where $U$ is  the unitary transformation  $\nu_\alpha  = U_{\alpha i}  \nu_i$
that puts the inverse neutrino propagator  in diagonal form. 
While the QKEs 
can be written in any basis, we give our results below  in the ``flavor'' basis.

\subsection{Green's function approach to the QKEs}

\subsubsection{Generalities}

The description in terms of creation and annihilation operators presented so far has a simple counterpart in the QFT approach 
of  Ref.~\cite{Vlasenko:2013fja}.   In that approach, the basic dynamical objects are the neutrino two-point functions ($a$ and $b$ denote 
flavor indices, and we suppress spinor indices)
\begin{subequations}
\begin{align}
\left(G^{(\nu)}_{ab} \right)^+ (x,y) & \equiv      \langle    \nu_a (x)  \bar{\nu}_b (y)  \rangle ~,
\\ 
\left(G^{(\nu)}_{ab} \right)^- (x,y)  & \equiv      \langle    \bar{\nu}_b (y)  \nu_a (x)   \rangle ~,
\end{align}
\end{subequations}
from which one can construct the statistical ($F$) and spectral ($\rho$) functions, 
\begin{subequations}
\begin{align}
F^{(\nu)}_{ab} (x,y)  & \equiv   \frac{1}{2}  \,  \langle  \, [\nu_a (x) , \bar{\nu}_b (y) ]  \, \rangle = \frac{1}{2}  
\left( \left(G^{(\nu)}_{ab} \right)^+ (x,y)  -  \left(G^{(\nu)}_{ab} \right)^- (x,y)  \right)
\,,\\
\rho^{(\nu)}_{ab} (x,y)  & \equiv  i \,  \langle  \{ \nu_a (x) , \bar{\nu}_b (y) \}  \rangle = i    
\left( \left(G^{(\nu)}_{ab} \right)^+ (x,y)  + \left(G^{(\nu)}_{ab} \right)^- (x,y)  \right)
\,,
\end{align}
\end{subequations}
and the time-ordered propagator
\be
G^{(\nu)}_{ab}  (x,y) \equiv      \langle   T \left(  \nu_a (x)  \bar{\nu}_b (y)  \right)  \rangle = 
\theta (x^0 - y^0)  \left(G^{(\nu)}_{ab} \right)^+ (x,y)  -  \theta (y^0 - x^0)  \left(G^{(\nu)}_{ab} \right)^- (x,y) ~.
\ee
The statistical and spectral function have a simple physical interpretation (see for example \cite{Berges:2004yj}): 
roughly speaking the spectral function encodes information on the spectrum of the theory, i.e. the states that are available, 
while the statistical function gives information about the occupation numbers and quantum coherence for the available states. 
As we will show below,  the Wigner Transform 
(i.e. Fourier transform with respect to the relative coordinate)  of the statistical function  
\be
F^{(\nu)}_{ab} (k, x)  =   \int \  d^4r  \  e^{i k \cdot r}  \ F^{(\nu)}_{ab} (x+r/2, x-r/2)
\ee
contains all the information about the density matrices introduced in Eqs~(\ref{eq:f1}) and (\ref{eq:FFbar1}).
Below we sketch the various steps leading to the QKEs.

\subsubsection{Equations of motion}

\begin{figure}[t]
\centering
\includegraphics[width=3.5in]{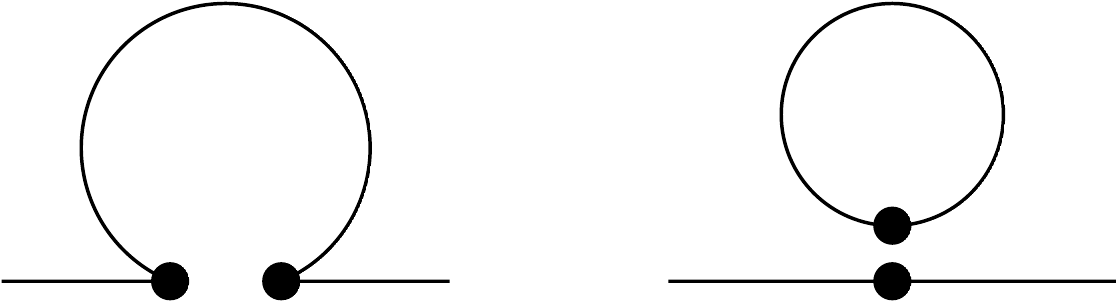}
\caption{Feynman graphs contributing to $\Sigma (x)$. 
External lines represent neutrinos. 
Internal  lines  represent $\nu, e, n, p$  propagators. We represent each  4-fermion interaction vertex  from 
Eqs.~(\ref{eq:interaction-lagrangian})  in terms of two displaced  fermionic current vertices.}
\label{fig:feynman1}
\end{figure}

Starting point is the equation of motion for the two point function $G^{(\nu)}_{ab}  (x,y)$, equivalent to the Dyson-Schwinger equation
\begin{align}
\left[ G^{(\nu)}  (x,y) \right]^{-1} =  \left[ G_0^{(\nu)}  (x,y) \right]^{-1}     -  \tilde{\Sigma} (x,y) ~, 
\end{align}
where we have suppressed for simplicity the flavor indices.  $G_0^{(\nu)}  (x,y)$ is the tree-level two point function and 
$\tilde{\Sigma} (x,y)$ is the neutrino self-energy, i.e. the sum of all  amputated one-particle-irreducible (1PI) diagrams with 
two external neutrino lines.   $\tilde{\Sigma} (x,y)$ is itself a functional of the two point function $G^{(\nu)}$ and admits the decomposition 
into a local term, and $^\pm$ components:
\begin{align}
\tilde{\Sigma} (x,y)=  -  i \Sigma (x)  \delta^{(4)} (x-y)   \ + \   \theta(x^0-y^0)  \tilde \Pi^+ (x,y)  \ - \  \theta(y^0-x^0)  \tilde  \Pi^- (x,y)~.
\end{align}
With the interactions given in Sect.~\ref{sect:int}, one can show that  $\Sigma (x)$ receives contributions starting at one loop (Fig.~\ref{fig:feynman1}), i.e. first order in $G_F$, 
while $\Pi^\pm (x,y)$ receive contributions starting at two loops (Fig.~\ref{fig:feynman2}), and are thus of second order in $G_F$. 

Wigner-transforming  the equation of motion for the two-point function and  keeping terms up to $O(\epsilon^2)$ 
in the small ratios  discussed in Sect.~\ref{sect:intro}, namely, 
\begin{align}
\frac{\partial_x, m, \Sigma}{E_\nu} &= O(\epsilon) \,, &
\frac{\tilde \Pi^\pm}{E} &= O(\epsilon^2)~, 
\end{align}
one arrives at~\cite{Vlasenko:2013fja}
\begin{subequations}
\begin{align}
\hat{\Omega} \, F^{(\nu)} (k,x)   &=   -\frac{i}{2}   \left(  \tilde \Pi^+ (k,x) G^{(\nu)-} (k,x)  - \tilde \Pi^-(k,x) G^{(\nu)+} (k,x) \right)
\,,\\
\hat{\Omega} &= \slashed{k} + \frac{i}{2}  \slashed{\partial}  - \Sigma (x) - m      + \frac{i}{2} \frac{\partial \Sigma}{\partial x^\mu} \frac{ \partial}{\partial k_\mu} ~.
\end{align}
\label{eq:proto-qke}
\end{subequations}

\subsubsection{Decomposition in spinor components}
 
The Wigner transform of the statistical function  $F^{(\nu)}_{ab} (k, x)$  (and any other two-point function) has sixteen spinor components (scalar, pseudoscalar, 
vector, axial-vector, tensor), 
\begin{align}
F^{(\nu)}  = 
\left[  F_S + \left( F_V^R\right)^\mu  \gamma_\mu   - \frac{i}{4}   \left( F_T^L \right)^{\mu \nu} \sigma_{\mu \nu} \right]  P_L + 
     \left[ F_S^\dagger+  \left( F_V^L \right)^\mu  \gamma_\mu  +  \frac{i}{4}   \left( F_T^R \right)^{\mu \nu} \sigma_{\mu \nu} \right]  P_R~, 
\end{align}
where $P_{L,R} \equiv (1 \mp \gamma_5)/2$ and $\s_{\m\n}\equiv\frac i2[\g_\m,\g_\n]$.
The various components satisfy the hermiticity conditions  $F_V^{L\dagger} = F_V^L$, 
$F_V^{R\dagger} = F_V^R$, and $F_T^{L\dagger} = F_T^R$.  
The forward scattering potential $\Sigma (x)$ and the inelastic collision self-energies $\tilde \Pi^\pm (k,x)$ admit a similar decomposition in spinor components
 (we give here only the decomposition for $\tilde{\Pi}$, a completely analogous one exists for $\Sigma$): 
\begin{align}
\tilde \Pi = \left[ \Pi_S  + \Pi_R^\mu  \gamma_\mu   - \frac{i}{4}   \left( \Pi_T^L \right)^{\mu \nu} \sigma_{\mu \nu} \right]  P_L + 
     \left[ \Pi_S^\dagger   + \Pi_L^\mu  \gamma_\mu  +  \frac{i}{4}   \left( \Pi_T^R \right)^{\mu \nu} \sigma_{\mu \nu} \right]  P_R~,
\end{align}
where we suppress the $^\pm$ superscripts.

For ultra-relativistic neutrinos  of three-momentum $\vec{k}$ (characterized by polar angle $\theta$ and azimuthal angle $\varphi$), it is convenient to express 
all Lorentz tensors and components of the two-point functions  (such as the $(F_V^{L,R})^\mu$ and $(F_T^{L,R})^{\mu \nu}$) in terms of a basis formed by 
two light-like four-vectors  $\hat\kappa^\mu (k) = (\sgn(k^0), \hat{k})$ and $\hat\kappa'^\mu (k) = (\sgn(k^0), -\hat{k})$
($\hat\kappa\cdot \hat\kappa = \hat\kappa' \cdot \hat\kappa' = 0$,  $ \hat\kappa \cdot \hat\kappa' = 2$) and 
two transverse four vectors  $\hat x_{1,2} (k)$ such that $\hat\kappa \cdot\hat x_i = \hat\kappa' \cdot \hat x_i = 0$ 
and $\hat x_i \cdot\hat x_j = - \delta_{ij}$, or equivalently  $\hat x^\pm \equiv \hat x_1 \pm i\hat x_2$, with  $\hat x^+\!\cdot\hat x^-=-2$ 
(see Appendix~\ref{app:kin} for additional details). 

The components of the self-energy entering the QKEs  are obtained by the  projections (see Appendix~\ref{app:kin})
\begin{subequations}\label{eq:projections-general}
\begin{align}
\Pi_{L,R}^\kappa (k,x) &= \frac{1}{2}  \hat{\kappa}_\mu  \, \Tr  \left[ \tilde \Pi  (k,x)\, \gamma^\mu P_{L,R} \right]
\,,\\
P_T (k,x)  &=  \frac{ie^{i\vp}}{16} \, (\hat \kappa \wedge\hat x^+)^{\mu \nu}  \,  \, \Tr  \left[ \tilde \Pi (k,x) \, \sigma_{\mu \nu}  P_{R} \right] ~, 
\qquad \quad  (a \wedge b)^{\mu \nu} = a^\mu b^\nu - a^\nu b^\mu 
\,,\\
\Sigma_{L,R}^\mu (x)  &= \frac{1}{2}  {\rm Tr}  \Big[   \Sigma (x) \gamma^\mu  P_{L,R}   \Big]
\,,
\end{align}
\end{subequations}
which can be arranged in the $2 n_f \times 2 n_f$ structures: 
\begin{align}
\Sigma^\mu (x)  &=
\left(
\begin{array}{cc}
\Sigma_R^\mu  & 0 \\
0 &  \Sigma_L^\mu 
\end{array}
\right)
\,, &
\hat \Pi^\pm (k,x)  &=
\left(\begin{array}{cc}
\Pi_R^{\kappa \pm} & 2P_T^\pm \\
2P_T^{\pm \dagger}  &  \Pi_L^{\kappa \pm} 
\end{array}
\right)~.
\label{eq:def-Pi-matrix}
\end{align}
%

\subsubsection{Leading order analysis}

The equations of motion (\ref{eq:proto-qke})  impose relations between the  sixteen components of $F^{(\nu)}_{ab} (k, x)$. 
Solving  (\ref{eq:proto-qke})  to $O(\epsilon^0)$,  only four components survive (L- and R- vector components and two tensor components), parameterized  
by the real functions $F_{L,R} (k,x)$ and the complex function $\Phi (k,x)$:
\begin{subequations}
\begin{align}
\left(F_V^{L,R} \right)^\mu \!(k,x) &=  \hat{\kappa}^\mu (k)  \,  F_{L,R}  (k,x) ~,
\\
\left(F_T^L \right)_{\mu \nu} \!(k,x)  &= e^{-i\vp (k)} (\hat{\kappa}  (k) \wedge \hat{x}^{-} (k))_{\mu \nu}   \, \Phi (k,x)~, 
\\
\left(F_T^R \right)_{\mu \nu}\!(k,x) &= e^{i\vp(k)} (\hat{\kappa} (k) \wedge \hat{x}^{+}(k))_{\mu \nu} \,  \Phi^\dagger (k,x)
~,
\end{align}
\end{subequations}
with all other spinor components vanishing. 

Beyond $O(\epsilon^0)$,  the four independent spinor components of $F^{(\nu)}_{ab} (k, x)$  can be conveniently chosen 
to coincide with the ones non-vanishing to $O(\epsilon^0)$~\cite{Vlasenko:2013fja,Cirigliano:2014aoa}. 
They can be  isolated by the following projections:
\begin{subequations}
\begin{align}
F_{L,R} (k,x) &\equiv  \frac{1}{4} {\rm Tr} \Big( \gamma_\mu  P_{L,R} \ F^{(\nu)} (k,x) \Big)  \, \hat\kappa'^{\mu} (k)
\,,\\
\Phi^{(\dagger)} (k,x) &\equiv   \mp \frac{i}{16}  {\Tr} \Big( \sigma_{\mu \nu}  P_{L/R}\ F^{(\nu)}(k,x) \Big)   (\hat\kappa' (k)  \wedge\hat x^\pm (k))^{\m\n}
e^{\pm i\vp (k)}
\,,
\label{eq:phi}
\end{align}
\end{subequations}
where the upper (lower) signs and indices refer to $\Phi$ ($\Phi^\dagger$) 
~\footnote{Compared to Ref.~\cite{Vlasenko:2013fja},  in Eq.~(\ref{eq:phi}) the appearance of the additional phase $\vp (k)$  (azimuthal angle of $\vec{k}$)  is due to our choice of coordinates \eqref{eq:basis-in-spherical}, as was explained in Ref.~\cite{Cirigliano:2014aoa}.}.
These components can be collected in a  $2 n_f \times 2 n_f$ matrix~\footnote{
Notice that the arrangement of the $L$, $R$ components here differs from the one in 
the matrix $\hat\Pi$ in \eqnref{eq:def-Pi-matrix} above.   
The  reason for this is that the Lorentz scalar components 
of the equations of motion \eqref{eq:proto-qke} (ultimately determining the kinetic equations for 
$F_L$ and $F_R$)  involve  derivatives acting on $F_{L,R}$   as well as  products of the type
$F_{L,R} \cdot \Sigma_{R,L}$ and $F_{L,R}  \cdot  \Pi^\kappa_{R,L}$. 
This can be verified either by direct matrix multiplication~\cite{Vlasenko:2013fja}  or by taking the appropriate traces.  
Therefore,  the ``R'' chiral components of the self-energy  affect the dynamics of the ``L'' density $F_L$,  and vice versa. 
With the choice made in Eqs.~(\ref{eq:def-Pi-matrix})  and (\ref{eq:Fhat0}) one then obtains 
kinetic and shell conditions in the compact  $2 n_f \times 2 n_f$ matrix form given below in \eqref{eq:qkec0} and \eqref{eq:shell}.
}
\be
\hat{F} = 
\left(
\begin{array}{cc}
F_L &  \Phi \\
\Phi^\dagger &  F_R
\end{array}
\right)~.
\label{eq:Fhat0}
\ee
In the free theory, the positive and negative frequency integrals of ${\hat F}$ 
give (up to a constant) the particle and antiparticle density matrices  of Eq.~(\ref{eq:FFbar1})
defined  in terms of ensemble averages of  creation and annihilation operators (see Eqs~(\ref{eq:f1})):
\begin{subequations}
\begin{align}
\label{eq:moment}
- 2 \int_0^\infty  \frac{dk^0}{2 \pi}  {\hat F} (k,x)  &=  F ( \vec{k},x)  - \frac{1}{2} \id
\,,\\
\label{eq:momentbar}
- 2 \int_{-\infty}^0  \frac{dk^0}{2 \pi}  {\hat F} (k,x)  &=  \bar{F}  (- \vec{k},x) - \frac{1}{2} \id~.
\end{align}
\end{subequations}
In the interacting theory, we take the above equations as definitions of neutrino and antineutrino  
number densities (generalized to include the off-diagonal coherence terms). 
These correspond to spectrally-dressed densities in the language adopted in Refs.~\cite{Dev:2014laa,Millington:2012pf}, 
with quasi-particle spectra dictated by Eq.~(\ref{eq:shell}) below.

\subsubsection{Kinetic  equations and  shell   conditions beyond leading order}
Beyond $O(\epsilon^0)$, the dynamics of  $\hat F(k,x)$ is still controlled by Eq.~(\ref{eq:proto-qke}). 
Projecting out all the  spinor components of (\ref{eq:proto-qke}) one finds~\cite{Vlasenko:2013fja}:
(i) constraint relations  that express ``small components'' of $F^{(\nu)}$ in terms of $F_{L,R}$ and $\Phi$; 
(ii)   evolution equations (i.e. first order in space-time derivatives) for $\hat F (k,x)$; 
(iii) constraint relations on the  components $\hat F(k,x)$, which determine the shell structure of the solutions. 
Defining  $\partial^\kappa \equiv  \hat\kappa (k) \cdot  \partial$,  $\partial^i \equiv\hat x^i  (k) \cdot \partial$, 
the kinetic and constraint equations  (items (ii) and (iii) above) for  the matrix $\hat F(k,x)$ are
\begin{subequations}
\be
 \partial^\kappa \hat F + \frac{1}{2 |\vec{k}|}  \left\{ \Sigma^i , \partial^i  \hat F \right\}  +  \frac{1}{2}  \left\{  \frac{\partial \Sigma^\kappa}{\partial x^\mu} , 
 \frac{\partial \hat  F}{\partial k_\mu }       \right\}    
= - i \left[ H, \hat{F} \right]   +   \hat C  
~,\label{eq:qkec0}
\ee
\be
\label{eq:shell}
\left\{  \hat {\kappa} (k) \cdot k - \Sigma^\kappa \ ,  \ \hat{F} \right\} = 0~. 
\ee
\end{subequations}
Using   $k^\mu = |\vec{k}| \,  \hat{\kappa}^\mu (k) +  O(\epsilon^2)$  the  constraint equation
(\ref{eq:shell}) can be written in the more familiar form of a shell-condition: 
\be
\left\{   k^2 - | \vec{k} | \, \Sigma^\kappa  (x) \ ,  \ \hat{F}  (k,x)  \right\} = 0~.
\ee
The $2 n_f \times 2 n_f$ potential $\Sigma (x)$   is defined in~(\ref{eq:def-Pi-matrix}) and its projections are $\Sigma^{\kappa} =  \hat \kappa (k) \cdot \Sigma$ and 
$\Sigma^i = \hat{x}_i \cdot \Sigma$. 

The Hamiltonian-like operator controlling the coherent evolution in Eq.~(\ref{eq:qkec0})  is  given by 
\be
H  =
\left(
\begin{array}{cc}
H_R &  H_{LR}  \\
H_{LR}^\dagger   & H_L
\end{array}
\right)~ \qquad 
\ee
with 
\begin{subequations}
\label{eq:hs}
\begin{align}
H_R  &=  \Sigma_R^\kappa   + \frac{1}{2 |\vec{k}|}   \left( m^\dagger m  - \epsilon^{ij} \partial^i \Sigma_R^j   + 4 \Sigma_R^+ \Sigma_R^-\right)
\,,\label{eq:HR}
 \\
H_L  &=  \Sigma_L^\kappa   + \frac{1}{2 |\vec{k}|}   \left( m  m^\dagger  + \epsilon^{ij} \partial^i \Sigma_L^j   + 4 \Sigma_L^- \Sigma_L^+\right)
\,,\label{eq:HL}
\\
H_{LR}  &=   - \frac{1}{|\vec{k}|}  \left( \Sigma_R^+  \, m^\dagger - m^\dagger \, \Sigma_L^+ \right) ~, 
\label{eq:Hm}
\end{align}
\end{subequations}
where $\Sigma_{L,R}^\pm \equiv (1/2) \,  e^{\pm i \varphi} \,   ( x_1 \pm i x_2)_\mu  \,   \Sigma^\mu_{L,R}$ and 
$\epsilon^{ij}$ is the two-dimensional  Levi-Civita symbol ($\epsilon^{12} = 1$).

Finally, the   collision term  in Eq.~(\ref{eq:qkec0})  reads
\be
\label{eq:Chat}
\hat{C}  = -  \frac{1}{2}  \left\{  \hat \Pi^+ , {\hat G}^- \right\}  + \frac{1}{2} \left\{ \hat \Pi^-,  {\hat G}^+ \right\}
\,, \\
\ee 
where the $2 n_f  \times 2 n_f$  gain and loss potentials  $ \hat \Pi^\pm (k)$   
are given in Eq.~(\ref{eq:def-Pi-matrix}) in terms of spinor components of the  self-energy, 
extracted from a calculation of the two-loop diagrams of Fig.~\ref{fig:feynman2}, 
and 
\be
\hat{G}^\pm  =  - \frac{i}{2}  \hat \rho \, \id   \pm \hat F ~. 
\ee
In order to obtain the collision terms to $O(\epsilon^2)$ we will need only the $O(\epsilon^0)$ expression for 
the vector component of the spectral function, namely  $\hat \rho (k)  = 2 i \pi  |\vec{k}|  \delta(k^2) {\rm sgn} (k^0)$ (see Appendix~\ref{sec:greenfcts}).

\subsubsection{Integration over frequencies: QKEs for Dirac and Majorana neutrinos}

The final step to obtain the QKEs requires integrating (\ref{eq:qkec0}) over positive and negative frequencies,  
taking into account the $O(\epsilon)$ shell corrections from (\ref{eq:shell}), whenever required  in order to 
keep terms up to $O(\epsilon^2)$ in power-counting. 
Recalling the definitions  (\ref{eq:moment}), the integrations over positive and negative frequency lead to:
\begin{subequations}
\label{eq:qkec1}
\begin{align}
 \partial^\kappa F  + \frac{1}{2 |\vec{k}|}  \left\{ \Sigma^i , \partial^i F \right\}  - \frac{1}{2}  \left\{  \frac{\partial \Sigma^\kappa}{\partial \vec{x}},  \frac{\partial F}{\partial \vec{k}}  \right\}      &= -i  [H,F]  +   {\cal C}~,
\\
\partial^\kappa \bar F  -  \frac{1}{2 |\vec{k}|}  \left\{ \Sigma^i , \partial^i \bar{F} \right\}  + \frac{1}{2}  \left\{  \frac{\partial \Sigma^\kappa}{\partial \vec{x}},  \frac{\partial \bar F}{\partial \vec{k}}       \right\}    &= - i [ \bar H, \bar F]  +   \bar{\cal C}~. 
\end{align}
\end{subequations}
The  differential operator on the left-hand side  generalizes the ``Vlasov'' term.
The first term on the right-hand side controls 
coherent evolution due to mass and forward scattering 
and generalizes the standard Mikheyev-Smirnov-Wolfenstein
(MSW) effect~\cite{Wolfenstein:1978,Mikheyev:1985,Balantekin:2007kx}. 
Finally, the second term  on the right hand side encodes inelastic collisions and generalizes the 
standard Boltzmann collision term~\cite{Keil:2003qy,Mezzacappa:2005lr,Kotake:2006fk,Brandt:2011mz,Ellinger:2013gf,
Cherry:2012lu,
Sarikas:2012fk,
Mirizzi:2012qy,
Cherry:2013lr}.
Let us now discuss in greater detail each term.

The physical meaning of the differential operators on the LHS of (\ref{eq:qkec1})  becomes more transparent by 
noting that they can be re-written as 
\be
\partial_t +  \frac{1}{2} \{ \partial_{\vec{k}}  \omega_\pm ,    \partial_{\vec{x}}  \  \}  - \frac{1}{2}  \{  \partial_{\vec{x}} \omega_\pm  , \partial_{\vec{k}} \   \}, 
\ee
with 
$\omega_+ =   |\vec{k}|  +  \Sigma^\kappa$ 
 for neutrinos and $\omega_- =  |\vec{k}|  -  \Sigma^\kappa $ for antineutrinos.
Recalling  that   $\omega_{\pm}  (\vec{k})   =  |\vec{k}|  \pm \Sigma^\kappa$ 
are the $O(\epsilon)$    neutrino ($+$) and antineutrino ($-$) Hamiltonian operators, 
one sees that  the  differential operators on the LHS of  (\ref{eq:qkec1}) 
generalize the total time-derivative operator
$d_t =  \partial_t   + \dot{\vec{x}}  \ \partial_{\vec{x}} +   \dot{\vec{k}}   \ \partial_{\vec{k}} $, 
with 
$\dot{\vec{k}}  = - \partial_{\vec{x}} \, \omega$
 and   
$\dot{\vec{x}}  =  \partial_{\vec{k}} \, \omega$, 
thus encoding  the familiar drift and force terms.

In terms of the mass matrix $m$ and the potentials $\Sigma^\mu_{L,R}$, 
the Hamiltonian-like operators controlling the coherent evolution are given by 
\begin{align}
H &=
\left(
\begin{array}{cc}
H_R &  H_{LR}  \\
H_{LR}^\dagger   & H_L
\end{array}
\right) ~, &
\bar H &=
\left(
\begin{array}{cc}
\bar{H}_R &  H_{LR}  \\
H_{LR}^\dagger   & \bar{H}_L
\end{array}
\right) ~, 
\end{align}
with  $H_L$, $H_R$, $H_{LR}$ given in Eqs.~(\ref{eq:hs}). 
The antineutrino operators $\bar{H}_{L,R}$ can be obtained from $H_{L,R}$ by flipping the sign of the entire term multiplying $1/(2 |\vec{k}|)$.  
The first two terms in $H_{L,R}$ are included in all analyses of neutrino oscillations in medium. 
$\Sigma^\kappa_{L,R}$ include the usual forward scattering off matter and neutrinos, and are functions of $F, \bar F$ 
thereby introducing  non-linear effects  in the coherent evolution. 
The $m^\dagger m/|\vec{k}|$ term encodes vacuum oscillations. 
The additional terms in $H_{L,R}$ and the spin-flip  term $H_{LR}$, 
discussed in detail in Refs.~\cite{Cirigliano:2014aoa,Vlasenko:2014bva},
complete the set of contributions to $O(\epsilon^2)$.

Finally, the   collision terms  on the RHS of Eqs.~(\ref{eq:qkec1})  are 
\begin{subequations}\label{eq:C-Cbar}
\begin{align}
{\cal C}  &= \frac{1}{2}  \left\{  \Pi^+ , {F} \right\}  - \frac{1}{2} \left\{ \Pi^-,  \id - {F} \right\}
\,, \\
\bar{\cal C} &= \frac{1}{2}  \left\{  \bar{\Pi}^+ , \bar{F} \right\}  - \frac{1}{2} \left\{ \bar{\Pi}^-,  \id - \bar{F} \right\} ~, 
\end{align}
\end{subequations}
where~\footnote{From now on we suppress the explicit  $x$-dependence in all Wigner transforms.}
\begin{align}
\Pi^\pm  (\vec{k}) &=  \int_0^\infty   dk^0  \   \hat \Pi^\pm (k^0,\vec{k}) \, \delta (k^0 - \abs{\vec{k}}) 
\,, &
\bar{\Pi}^\pm   (\vec{k}) &=  - \int_{-\infty}^0   dk^0  \  \hat \Pi^\mp (k^0,- \vec{k}) \, \delta (k^0 +\abs{\vec{k}})
\,. \label{eq:Pi-Pibar}
\end{align}
and the  $2 n_f  \times 2 n_f$  gain and loss potentials $\hat\Pi^\pm$  
are given in Eq.~(\ref{eq:def-Pi-matrix}) in terms of spinor components of the  self-energy. 

The above discussion directly applies to Dirac neutrinos, with 
Eqs.~(\ref{eq:qkec1}) representing QKEs for neutrino ($F(k,x)$)  
and antineutrino ($\bar F(k,x)$) density matrices.  
In the Majorana case $\bar F(k,x)$ contains no additional information compared to $F(k,x)$. 
So one can get QKEs for Majorana neutrino exclusively from the positive frequency integral 
of (\ref{eq:qkec0}).  To avoid confusion,  in the Majorana case we denote the 
positive frequency integral of $\hat{F} (k,x)$  by ${\cal F} (k,x)$  (see  Eq.~(\ref{eq:FMajorana}) for a discussion 
of its physical content).  
The Majorana QKE is formally identical to the first one of (\ref{eq:qkec1}):
\begin{align}
 \partial^\kappa {\cal F}  + \frac{1}{2 |\vec{k}|}  \left\{ \Sigma^i , \partial^i {\cal F} \right\}  - \frac{1}{2}  \left\{  \frac{\partial \Sigma^\kappa}{\partial \vec{x}},  \frac{\partial {\cal F}}{\partial \vec{k}}  \right\}  &= -i  [H, {\cal F}]  +   {\cal C}_M~,
\label{eq:qkem1}
\end{align}
with
\begin{align}
{\cal C}_M  &= \frac{1}{2}  \left\{  \Pi^+ , {\cal F} \right\}  - \frac{1}{2} \left\{ \Pi^-,  \id - {\cal F} \right\}~,
\end{align}
and $\Pi^\pm$ formally given in (\ref{eq:Pi-Pibar}).
The analogy, however, is only superficial because the potentials $\Sigma_\mu$ and $\Pi^{\pm}$ have a very different structure 
in the Dirac and Majorana cases.  Anticipating results to be described later, 
we note that:
\begin{itemize}
\item Concerning the potentials induced by forward scattering, $\Sigma_{R,L}$,
for Dirac neutrinos $\Sigma_R \neq 0$ while $\Sigma_L \propto G_F  m^2 \sim O (\epsilon^3) $ (massless right-handed neutrinos do not interact). 
On the other hand, in the Majorana case  one has $\Sigma_L = - \Sigma_R^T$, with transposition acting on flavor indices: 
right-handed antineutrinos do interact  even in the massless limit. 

\item Concerning the inelastic potentials  $\hat \Pi^\pm$ given in (\ref{eq:def-Pi-matrix}),  
in the Majorana case all four $n_f \times n_f$ blocks are non-vanishing, i.e.  $\Pi_{R,L}^{\kappa \pm} \neq 0$ and  $P_T^\pm \neq 0$. 
In the Dirac case, on the other hand,  in the massless limit only  the upper diagonal block is non-vanishing, 
i.e.  $\Pi_{R}^{\kappa \pm} \neq 0$ and    $\Pi_{L}^{\kappa \pm} =  P_T^\pm = 0$. 
This again corresponds to the fact that massless right-handed neutrinos and left-handed antineutrinos
have no interactions in the Standard Model. 

\end{itemize}

\subsection{Refractive effects}

Before discussing in detail the collision terms in the next section,  for   completeness we
briefly describe refractive effects in the coherent evolution,  controlled by the 
potential $\Sigma$. 
$\Sigma_R$  and  $\Sigma_L$ (see Eq.~\ref{eq:def-Pi-matrix})  are the
4-vector potentials induced by forward scattering
for left-handed and right-handed neutrinos, respectively.
For Dirac neutrinos $\Sigma_R \neq 0$  and  $\Sigma_L \propto G_F  m^2 \sim O (\epsilon^3) $
while for Majorana  neutrinos   $\Sigma_L = - \Sigma_R^T$.

The potential induced by a  background of electrons and positrons is given  for any geometry by the following 
expressions:   
\begin{subequations}
\begin{align}
\left[ \Sigma_R^\mu  \ \Big \vert_{e}  
\right]_{ab}
&=
2\sqrt{2} G_F   \, 
\left[  \left(  \delta_{ea} \delta_{eb}  + \ \delta_{ab}  \, \left(\sin^2\theta_W-\frac{1}{2}\right)\right)J_{(e_L)}^\mu+ 
\delta_{ab} \ \sin^2\theta_W \,  J_{(e_R)}^\mu \right]
\,, \\
J^\mu_{(e_L)}   (x) &= \int \frac{d ^3q}{(2 \pi)^3}   \  v_{(e)}^\mu (q) \  \Big( f_{e_L} (\vec{q},x) - \bar{f}_{e_R} (\vec{q},x) \Big)~,
\\
J^\mu_{(e_R)}   (x) &= \int \frac{d ^3q}{(2 \pi)^3}   \  v_{(e)}^\mu (q) \  \Big( f_{e_R} (\vec{q},x) - \bar{f}_{e_L} (\vec{q},x) \Big)~,
\end{align}
\end{subequations}
where  $a,b$ are flavor indices, $v_{(e)}^\mu =    (1, \vec{q} / \sqrt{m_e^2 + q^2} \,  )$,   and we use the notation $f_{e_L} (\vec{q},x)$  ($\bar{f}_{e_L} (\vec{q},x)$)
for  the distribution function of  L-handed electrons (positrons), etc. 

The nucleon-induced potentials have similar expressions, with appropriate 
replacements of the L- and R-handed couplings to the $Z$  and the 
distribution functions $f_{e_L} \to f_{N_L}$, etc. 
For unpolarized electron and nucleon backgrounds of course one has $f_{e_L} = f_{e_R}$, etc., 
and the nucleon contribution to the potential is:
\be
\left[ \Sigma_R^\mu  \ \Big \vert_{N}  
\right]_{ab}
=  \sqrt{2} G_F   \,  C_V^{(N)}  \   J_{(N)}^\mu   \  \delta_{ab}~,
\ee
with $C_{V}^{(n,p)}$ given in Eq.~(\ref{eq:nucleoncouplings}).

Finally, the neutrino-induced potentials are given by
\begin{subequations}
\begin{align}
\label{eq:Sigma}
\left[ \Sigma_R^\mu  \ \Big \vert_{\nu} \right]_{ab}  &=
 \sqrt{2} G_F 
 \left(
\left[ J^{\mu}_{(\nu)}  \right]_{ab}  \ + \   \delta_{ab}   \  \Tr   J^\mu_{(\nu)}
 \right) ~,
\\
J^\mu_{(\nu)}  (x) &=  \int \frac{d ^3q}{(2 \pi)^3}   n^\mu (q) \  \Big( f_{LL} (\vec{q},x) - \bar{f}_{RR} (\vec{q},x) \Big)~,
\label{eq:Sigmav2}
\end{align}
\end{subequations}
with $n^\mu (q) = (1, \hat q)$. 
For a test-neutrino of three-momentum $\vec{k}$, these potentials  can be  further projected along the basis vectors: 
with light-like component $\Sigma^\kappa \equiv  n (k) \cdot \Sigma$ 
along the neutrino trajectory (in the massless limit); 
and space-like component $\Sigma^i \equiv x^i (k)  \cdot \Sigma$,  transverse to the neutrino trajectory.  
In particular, for the neutrino-induced contribution we find 
 $\Sigma^\kappa (x) \propto    \int d^3 q  \ (1 - \cos \theta_{k q} )  \cdot ( f_{LL} (\vec{q},x) - \bar{f}_{RR} (\vec{q},x))   $, consistently with the familiar results in the literature  (see \cite{Duan:2010fr} and references therein).

Having summarized the structure of the neutrino QKEs,   we next discuss the main new results of this paper, namely the 
calculation of the collision terms.

\begin{figure}
\centering
\includegraphics[width=5.in]{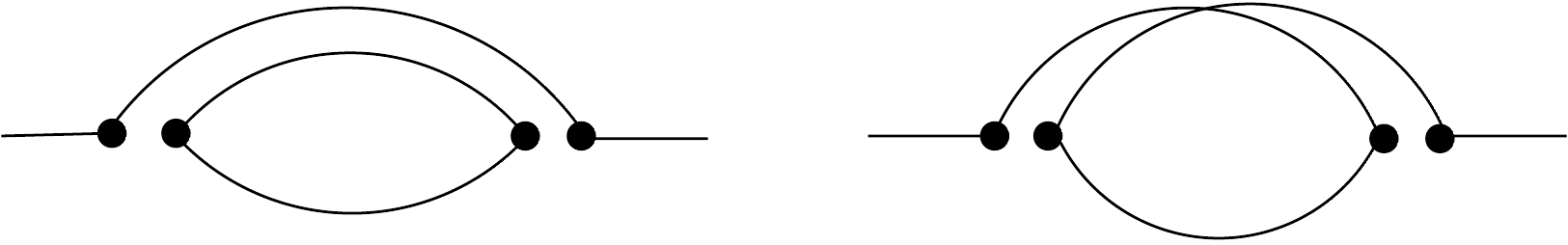}
\caption{Feynman graphs contributing to $\Pi (k,x)$. 
We represent each  4-fermion interaction vertex  from 
Eqs.~(\ref{eq:interaction-lagrangian})  in terms of two displaced  fermionic current vertices.
External lines represent neutrinos. 
Internal  lines  represent $\nu, e, n, p$  propagators (left diagram) and $\nu$ propagators (right diagram). 
}
\label{fig:feynman2}
\end{figure}

\section{Majorana collision term:  derivation and general  results}
\label{sect:derivation}
 
Direction-changing scattering  and 
inelastic processes such as neutrino emission, absorption, pair production, or pair annihilation 
are  encoded in ${\cal C}_M$ 
on the RHS of  \eqref{eq:qkem1}. 
In this section  we derive the structure and detailed  expressions for ${\cal C}_M$,  
providing, for the sake of completeness, several intermediate steps in the derivation.
All the results presented in this section pertain to Majorana neutrinos. 
The collision term for Dirac neutrinos is discussed in Section~\ref{app:dirac}.

We recall that the   Majorana  collision term  is given by 
\begin{subequations}\label{eq:C-CbarM}
\begin{align}
{\cal C}_M  &= \frac12  \left\{  \Pi^+ , {\cal F} \right\}  - \frac{1}{2} \left\{ \Pi^-,  \id - {\cal F} \right\}
\,, \\
\Pi^\pm  (\vec{k}) &= \int_0^\infty   dk^0  \  \hat \Pi^\pm (k^0,\vec{k}) \, \delta (k^0 - \abs{\vec{k}}) ~, 
\label{eq:Pi-Pibar-v2}
\end{align}
\end{subequations}
with the $2 n_f  \times 2 n_f$  gain and loss potentials $\Pi^\pm$  
given in Eq.~(\ref{eq:def-Pi-matrix}) in terms of spinor components of the  self-energy, 
extracted from a calculation of the two-loop self-energies 
of Fig.~\ref{fig:feynman2}.  
The $2 n_f  \times 2 n_f$ collision term matrix is thus given by 
\begin{align}
 {\cal C}_M=\begin{pmatrix}
    C & C_\phi \\
    C_\phi^\dagger & \bar{C}^T
   \end{pmatrix}
\label{eq:CMajorana}
\end{align}
with 
\begin{align}
 C(\vec{k})&=\frac12\left(\{(\Pi_{R}^{\kappa})^+,f\}-\{(\Pi_{R}^{\kappa})^-,(1\!-\!f)\}\right)
 +(P_T^+  + P_T^-)\phi^\dagger+\phi(P_T^+ + P_T^-)^\dagger 
 \,,\nn\\
\bar{C}^T(\vec{k})  &=
\frac12\left(\{(\Pi_{L}^{\kappa})^+, \bar{f}^T\}-\{(\Pi_{L}^{\kappa})^-,(1\!-\! \bar{f}^T)\}\right)
 +(P_T^+  + P_T^-)^\dagger \phi +\phi^\dagger  (P_T^+ + P_T^-) 
 \,,\nn\\
 C_\phi(\vec{k})&=\frac12\!\left(\!\big((\Pi_{R}^{\kappa})^+ +(\Pi_{R}^{\kappa})^-\big)\phi +\phi\big((\Pi_L^{\kappa})^+ +(\Pi_{L}^{\kappa})^-\big)\!\right)
+  f \, (P_T^+ + P_T^-)   + (P_T^+ + P_T^-)  \, \bar{f}^T  -  2 P_T^-
 \,. \label{eq:collisionterms}
\end{align}
Note that the collision term has a  non-diagonal matrix structure in both flavor~\cite{Sigl:1992fn,Raffelt:1992uj}
and spin space~\cite{Vlasenko:2013fja}.
The matrix components of $\Pi^\pm$  can be expressed in terms of neutrino density matrices and  distribution functions
of the medium particles  (electrons, etc.). 
To the order we are working, we need only the $O(\epsilon^0)$ expressions for the neutrino and matter Green's functions 
in the collision term.  These expressions are collected in Appendix~\ref{sec:greenfcts}.

The contribution to $\nu-\nu$ scattering neglecting spin coherence  is given in Ref.~\cite{Vlasenko:2013fja}.
We present below the full analysis  including scattering off neutrinos, electrons, and low-density nucleons.
Deriving the  collision term requires the following steps:
\begin{itemize}
\itemsep=3pt
\item Calculation of the self-energy  diagrams in  Fig.~\ref{fig:feynman2}.
\item Identification of the components of $\hat \Pi^\pm$  (see (\ref{eq:def-Pi-matrix}))
by projecting the self-energy diagrams  on appropriate spinor and Lorentz components via Eq.(\ref{eq:projections-general}). 
\item Integration over positive frequencies  according to  Eq.~(\ref{eq:Pi-Pibar-v2}) to obtain $\Pi^\pm (\vec{k})$.  
\item  Matrix multiplications to obtain the various components of  ${\cal C}_M$ in (\ref{eq:collisionterms}). 
\end{itemize}
In the following subsections we  present results on each of the above items.

\subsection{Self-energies to two loops}
\label{sect:diagrams}

The Standard Model interactions allow for neutrino-neutrino processes 
and neutrino interactions with charged leptons and nucleons, cf. \eqnref{eq:interaction-lagrangian}.
We report below the various contributions.

\begin{itemize}
\item Neutrino-neutrino processes receive contributions from both topologies in Fig.~\ref{fig:feynman2}.\\
Left diagram:
\begin{align}
\tilde{\Pi}_{ab}^\pm  (k)  &=  - G_F^2  \ \int\frac{d^4 q_1 \, d^4 q_2  \, d^4 q_3}{(2\pi)^8}\, \delta^{(4)} (k - q_3 -  q_1 +  q_2) 
\nn \\
&\quad \times \gamma_\mu (P_L-P_R)   \, G_{ab}^{(\nu) \pm} (q_3) \,  \gamma_\nu (P_L - P_R) \ 
\nn \\
&\quad \times  \Tr \left[
\gamma^\nu (P_L-P_R)   \, G_{cd}^{(\nu) \mp} (q_2)  \,  \gamma^\mu (P_L - P_R) \ G_{dc}^{(\nu) \pm} (q_1)
\right]
\label{eq:nunutrace}
\end{align}
Right diagram: 
\begin{align}
\tilde{\Pi}_{ab}^\pm  (k)  &=  2 G_F^2  \ \int  \frac{d^4 q_1 \, d^4 q_2  \, d^4 q_3}{(2\pi)^8}\, \delta^{(4)} (k - q_3 -  q_1 +  q_2) 
\nn \\
&\quad  \times \gamma_\mu (P_L-P_R)   \, G_{aa'}^{(\nu) \pm} (q_1) \,  \gamma_\nu (P_L - P_R) \, 
\nn \\
&\quad \times 
\, G_{a'b'}^{(\nu) \mp} (q_2)  \,  \gamma^\mu (P_L - P_R) \ G_{b'b}^{(\nu) \pm} (q_3)
\, \gamma^\nu (P_L-P_R)   
\qquad \qquad
\end{align}
where $a,b,c,d$ are flavor indices.

\item Neutrino-electron processes receive contributions only from the first topology in Fig.~\ref{fig:feynman2}:
\begin{align}
\tilde{\Pi}_{ab}^\pm  (k)  &=  - 8 G_F^2  \ \int  \frac{d^4 q_1 \, d^4 q_2  \, d^4 q_3}{(2\pi)^8}\, \delta^{(4)} (k - q_3 -  q_1 +  q_2) 
\nn \\
&\quad  \times  \sum_{A,B = L,R} \bigg\{
\gamma_\mu (P_L-P_R)   \,  \left[Y_{A}  G^{(\nu) \pm} (q_3)   Y_{B} \right]_{ab}  \,  \gamma_\nu (P_L - P_R) \ 
\nn \\
&\quad \times  \Tr \Big[
\gamma^\nu P_B  \, G^{(e) \mp} (q_2)  \,  \gamma^\mu P_A \ G^{(e) \pm} (q_1)
\Big] \bigg\}
\end{align}

\item Neutrino-nucleon processes receive contributions only from the first topology in Fig.~\ref{fig:feynman2}. 
There are two contributions, scattering and absorption. \\
Scattering: 
\begin{align}
\tilde{\Pi}_{ab}^\pm  (k)  &=  - 2 G_F^2  \ \int  \frac{d^4 q_1 \, d^4 q_2  \, d^4 q_3}{(2\pi)^8}\, \delta^{(4)} (k - q_3 -  q_1 +  q_2) 
\nn \\
&\quad  \times   \sum_{N = n,p} \bigg\{
\gamma_\mu (P_L-P_R)   \,   G_{ab} ^{(\nu) \pm} (q_3)    \,  \gamma_\nu (P_L - P_R) \ 
\   \Tr \Big[
\Gamma_N^\nu  \, G^{(N) \mp} (q_2)  \,  \Gamma_N^\mu  \ G^{(N) \pm} (q_1)
\Big] \bigg\}
\,, \nn\\
\Gamma_N^\mu &=  \gamma^\mu  \left(C_V^{(N)}  -  C_A^{(N)} \gamma_5\right)
\,.
\end{align}
Neutrino absorption and emission:
\begin{align}
\tilde{\Pi}_{ab}^\pm  (k)  &=  - 2 G_F^2 
 \ \int  \frac{d^4 q_1 \, d^4 q_2  \, d^4 q_3}{(2\pi)^8}\, \delta^{(4)} (k - q_3 -  q_1 +  q_2) 
\nn \\
&\qquad  \times  
\gamma_\mu P_L  \  \left[ I_e \right]_{ab}   \,  G^{(e) \pm} (q_3)    \  \gamma_\nu  P_L \ 
 {\rm Tr} \Big[ \Gamma^\mu  \, G^{(p) \pm} (q_1)  \,  \Gamma^\nu  \ G^{(n) \mp} (q_2)
\Big] 
\nn\\
&\quad - 2 G_F^2 
 \ \int  \frac{d^4 q_1 \, d^4 q_2  \, d^4 q_3}{(2\pi)^8}\, \delta^{(4)} (k + q_3 - q_1 +  q_2) 
\nn \\
&\qquad  \times 
\gamma_\mu P_R  \ \left[ I_e \right]_{ab}     G^{(e) \mp} (q_3)    \  \gamma_\nu  P_R \ 
{\rm Tr} \Big[ \Gamma^\mu  \, G^{(n) \pm} (q_1)  \,  \Gamma^\nu  \ G^{(p) \mp} (q_2)
\Big] 
\,, \nn\\
\Gamma^\mu &=  \gamma^\mu  \left(1  -  g_A  \gamma_5\right)
\,.
\label{eq:selfCC}
\end{align}
In the above expression we have introduced the  projector  on the electron flavor, 
\be
\left[ I_e \right]_{ab}    \equiv  \  \delta_{ae} \delta_{be} 
\label{eq:flprojector}
\ee
and we are neglecting contributions from $\mu^\pm$ and $\tau^\pm$, which are kinematically suppressed 
at the energies and temperatures of interest.  Their contributions are formally identical to the electron one, 
with the replacements $I_e \to I_{\mu}, I_\tau$ and $G^{(e)} \to G^{(\mu)}, G^{(\tau)}$.

\end{itemize}

\subsection{Projections on Lorentz structures}
Using the  identities collected in Appendix~\ref{app:kin2} and ~\ref{app:kin3}, 
one can perform all the needed projections in a straightforward way 
(at most four gamma matrices and a $\gamma_5$ appear in the traces).
In the following, we will suppress flavor indices.
The vector and tensor components of the self-energies (defined in \eqnref{eq:projections-general}) 
for all  processes considered in this work are:

\subsubsection*{Neutrino-nucleon scattering processes}

\begin{subequations}\label{eq:nucleon-scatt-proj}
\begin{align}
(\Pi_R^\kappa)^\pm (k) &= -\frac1{\abs{\vec{k}}} \int  \frac{d^4q_3}{(2 \pi)^4}
\left(R_N^\pm (k,q_3) \right)^{\mu \nu}  \ 
    \left[ \bar{G}_V^{L} (q_3)  \right]^{\pm}      \
  (T_R)_{\mu \nu} (k,q_3)
\,,\\
(\Pi_L^\kappa)^\pm (k) &= -\frac1{\abs{\vec{k}}} \int  \frac{d^4q_3}{(2 \pi)^4}
\left(R_N^\pm (k,q_3) \right)^{\mu \nu}  \ 
    \left[  \bar{G}_V^{R} (q_3)  \right]^{\pm}    \
  (T_L)_{\mu \nu} (k,q_3) 
 \,, \\
 P_T^\pm   (k) &=  \mp \frac{1}{2} \,        \int  \frac{d^4q_3}{(2 \pi)^4}  
\left(R_N^\pm (k,q_3) \right)^{\mu \nu}  \ 
\big[   \Phi (q_3)   \big]   \
  (T_T)_{\mu \nu} (k,q_3) ~. 
\end{align}
\end{subequations}
The various tensors are given by: 
\begin{subequations}
\label{eq:LRTtensors}
\begin{align}
(T_{R,L})_{\mu \nu} (k,k') &=   k_\mu k'_\nu  + k_\nu k'_\mu - k \cdot k'  \eta_{\mu \nu} \mp i \epsilon_{\mu \nu \alpha \beta} k^\alpha k'^\beta
\,, \\
(T_T)_{\mu \nu}  (k,k')&= [\hat{\kappa}(k')  \wedge \hat{x}^{-} (k')  ]_{\mu \alpha}  \ [ \hat{\kappa} (k)  \wedge \hat{x}^+  (k)] _{\nu}^{  \ \alpha}
 e^{i\left(\vp(k)-\vp(k')\right)} ~. 
\end{align}
\end{subequations}
In the last expression we have explicitly indicated  the dependence of  the basis vectors $\hat{\kappa},\hat{ \kappa}', \hat{x}_{\pm}$ 
on the four-momenta $k$ and $k'$. 
Finally,  the nucleon response function is given by:
\beq
\left(R_N^\pm (k,q_3) \right)^{\mu \nu}   =   2 G_F^2    \,    \int   \frac{d^4q_1}{(2 \pi)^4} \frac{d^4q_2}{(2 \pi)^4}   
\, (2\pi)^4  \delta^{(4)}  (k - q_3 - q_1  + q_2)
\   \Tr \Big[ \Gamma_N^\mu  \, G^{(N) \pm} (q_1) 
 \,  \Gamma_N^\nu    \, G^{(N) \mp} (q_2)  \Big]
 .
\eeq
We give the explicit expressions for the neutrino and nucleon Green's functions $\bar{G}_V^{L,R}$,  $\Phi$, and $G^{(N)}$ 
to  $O(\epsilon^0)$  in Appendix~\ref{sec:greenfcts}.

\subsubsection*{Neutrino-electron processes}

\begin{subequations}
\begin{align}
(\Pi_R^\kappa)^\pm   (k) &=
 -     \sum_{A,B = L,R}  
 \   \frac{1}{\abs{\vec{k}}}  \int  \frac{d^4q_3}{(2 \pi)^4}  
\left(R_{BA}^\pm (k,q_3) \right)^{\mu \nu}  \ 
\big[ Y_{A} \,   \left(\bar{G}_V^{L} (q_3)  \right)^{\pm}  Y_{B}  \big]    \
  (T_R)_{\mu \nu} (k,q_3)
\,,\\
(\Pi_L^\kappa)^\pm   (k) &= 
 -     \sum_{A,B = L,R}  
 \   \frac{1}{\abs{\vec{k}}} \int  \frac{d^4q_3}{(2 \pi)^4}  
\left(R_{BA}^\pm (k,q_3) \right)^{\mu \nu}  \ 
\big[ Y_{A} \,   \left(\bar{G}_V^{R} (q_3)  \right)^{\pm}  Y_{B}  \big]    \
  (T_L)_{\mu \nu} (k,q_3)
 \,, \\
 P_T^\pm   (k) &=  \mp \frac{1}{2} \, 
  \sum_{A,B = L,R}   \  
      \int  \frac{d^4q_3}{(2 \pi)^4}  
\left(R_{BA}^\pm (k,q_3) \right)^{\mu \nu}  \ 
\big[  Y_{A}  \Phi (q_3)  Y_{B} \big]   \
  (T_T)_{\mu \nu} (k,q_3) ~. 
\end{align}
\end{subequations}
The electron response functions are given by (recall $A,B  \in \{ L,R \}$ and $P_{A,B} = P_{L,R}$):
\beq
\left(R_{BA}^\pm (k,q_3) \right)^{\mu \nu}   =   8 G_F^2    \,       \int \frac{d^4q_1 d^4q_2}{(2 \pi)^4}
\, \delta^{(4)}  (k - q_3 - q_1  + q_2)
\   {\rm Tr}  \Big[ \gamma^\mu \, P_A \, G^{(e) \pm} (q_1) 
 \,  \gamma^\nu  \, P_B  \, G^{(e) \mp} (q_2)  \Big] .
\eeq
Expressions for the electron Green's functions $G^{(e)}$  
to  $O(\epsilon^0)$  are given in Appendix~\ref{sec:greenfcts}.

\subsubsection*{Charged-current processes}

\begin{subequations}
\begin{align}
(\Pi_R^\kappa)_{ab}^\pm   (k) &= - 
 \frac{1}{\abs{\vec{k}}}         \int  \frac{d^4q_3}{(2 \pi)^4}  
\left(R_{CC}^\pm (k,q_3) \right)^{\mu \nu}  \ 
    \left[      \left[ I_e \right]_{ab}  \, 
     \bar{G}^{(e) \pm}   (q_3)  \right]      \
  (T_R)_{\mu \nu} (k,q_3)  
 \,, \\
(\Pi_L^\kappa)_{ab}^\pm   (k) &= - 
 \frac{1}{\abs{\vec{k}}}         \int  \frac{d^4q_3}{(2 \pi)^4}  
\left(\tilde{R}_{CC}^\pm (k,q_3) \right)^{\mu \nu}  \ 
    \left[ 
   \left[ I_e \right]_{ab}  \,  
    \bar{G}^{(e) \mp}   (q_3)  \right]      \
  (T_L)_{\mu \nu} (k,q_3)
  \,.
\end{align}
\end{subequations}
There is no tensor projection from these processes. The charged-current response is given by:
\begin{subequations}
\begin{align}
\left(R_{CC}^\pm (k,q_3) \right)^{\mu \nu}   &=   2 G_F^2    \,     \int \frac{d^4q_1 d^4q_2}{(2 \pi)^4}
\, \delta^{(4)}  (k - q_3 - q_1  + q_2)
\Tr  \Big[ \Gamma^\mu  \, G^{(p) \pm} (q_1) 
 \,  \Gamma^\nu    \, G^{(n) \mp} (q_2)  \Big]
\,, \\
\left(\tilde{R}_{CC}^\pm (k,q_3) \right)^{\mu \nu}   &=   2 G_F^2    \,   \int \frac{d^4q_1 d^4q_2}{(2 \pi)^4}
\, \delta^{(4)}  (k + q_3 - q_1  + q_2)
\Tr  \Big[ \Gamma^\mu  \, G^{(n) \pm} (q_1) 
 \,  \Gamma^\nu    \, G^{(p) \mp} (q_2)  \Big]~.
\end{align}
\end{subequations}

\subsubsection*{Neutrino-neutrino processes}
The diagram  in the left panel of Fig.~\ref{fig:feynman2} induces:
\begin{subequations}
\begin{align}
(\Pi_R^\kappa)^\pm (k) &= -\frac{1}{\abs{\vec{k}}}  \int  \frac{d^4q_3}{(2 \pi)^4}
\left(R_{(\nu)}^\pm (k,q_3) \right)^{\mu \nu}  \ 
    \left[ \bar{G}_V^{L} (q_3)  \right]^{\pm}      \
  (T_R)_{\mu \nu} (k,q_3)
\,, \\
(\Pi_L^\kappa)^\pm (k) &= -\frac{1}{\abs{\vec{k}}}  \int  \frac{d^4q_3}{(2 \pi)^4}
\left(R_{(\nu)}^\pm (k,q_3) \right)^{\mu \nu}  \ 
    \left[  \bar{G}_V^{R} (q_3)  \right]^{\pm}    \
  (T_L)_{\mu \nu} (k,q_3)
 \,, \\
 P_T^\pm   (k) &=  \mp \frac{1}{2} \,        \int  \frac{d^4q_3}{(2 \pi)^4}  
\left(R_{(\nu)}^\pm (k,q_3) \right)^{\mu \nu}  \ 
\big[   \Phi (q_3)   \big]   \
  (T_T)_{\mu \nu} (k,q_3) ~. 
\\
\left(R_{(\nu)}^\pm (k,q_3) \right)_{\!\mu \nu}\!   &=  
2 \,  G_F^2    \,  \int \frac{d^4q_1 d^4q_2}{(2 \pi)^4}  
\,  \delta^{(4)}  (k - q_3 - q_1  + q_2)
\nn \\
&
\quad 
\times 
\Tr\Bigg[
\left[\bar{G}_V^L (q_2) \right]^\mp  \left[ \bar{G}_V^L   (q_1)   \right]^\pm  \,   (T_R)_{\mu \nu} (q_2,q_1) 
+ 
\left[\bar{G}_V^R (q_2) \right]^\mp  \left[  (\bar{G}_V^R (q_1)  \right]^\pm     \,   (T_L)_{\mu \nu} (q_2,q_1) 
\nonumber \\
&
\qquad\qquad 
- \Phi^\dagger (q_2) \Phi (q_1)  (T_T)_{\mu \nu} (q_2,q_1)  
- \Phi (q_2) \Phi^\dagger (q_1)  (T_T)^*_{\mu \nu} (q_2,q_1)  
\Bigg].
\end{align}
\end{subequations}

The diagram  in the right  panel of Fig.~\ref{fig:feynman2} induces 
\begin{subequations}
\begin{align}
(\Pi_R^\kappa)^\pm (k) &=  -  \frac{8 \, G_F^2  }{\abs{\vec{k}}}
 \int  \frac{d^4q_1 d^4q_2 d^4q_3}{(2 \pi)^8}  
\,  \delta^{(4)}  (k - q_3 - q_1  + q_2) 
\nn \\
&\quad \times  \bigg\{
k^\alpha q_2^\beta \,   (T_T)_{\alpha \beta} (q_3,q_1)  \ \Phi (q_1)  \,  \left[\bar{G}_V^{R} (q_2) \right]^\mp \, \Phi^\dagger (q_3) 
\nn \\
&\quad\qquad 
- k^\alpha q_3^\beta \,   (T_T)_{\alpha \beta} (q_2,q_1)  \ \Phi (q_1)  \, \Phi^\dagger (q_2) \,   \left[\bar{G}_V^{L} (q_3) \right]^\pm 
\nn \\
&\quad\qquad 
- k^\alpha q_1^\beta \,   (T_T)_{ \beta \alpha} (q_3,q_2)  \,   \left[\bar{G}_V^{L} (q_1) \right]^\pm    \,  \Phi (q_2)  \, \Phi^\dagger (q_3) 
\nn \\
&\quad\qquad 
+ 2 (k q_2)  (q_1 q_3)    \left[\bar{G}_L(q_1) \right]^\pm    \left[\bar{G}_L(q_2) \right]^\mp  \left[\bar{G}_L(q_3) \right]^\pm   
\bigg\}\,,
\\
(\Pi_L^\kappa)^\pm   (k) &= 
-  \frac{8 \, G_F^2  }{\abs{\vec{k}}}
 \int  \frac{d^4q_1 d^4q_2 d^4q_3}{(2 \pi)^8}  
\,  \delta^{(4)}  (k - q_3  - q_1 +q_2)
\nn \\
&\quad \times  \bigg\{
k^\alpha q_2^\beta \,   (T_T)^*_{\alpha \beta} (q_3,q_1)  \ \Phi^\dagger (q_1)  \,  \left[\bar{G}_V^{L} (q_2) \right]^\mp \, \Phi (q_3) 
\nn \\
&\quad\qquad 
- k^\alpha q_3^\beta \,   (T_T)^*_{\alpha \beta} (q_2,q_1)  \ \Phi^\dagger (q_1)  \, \Phi (q_2) \,   \left[\bar{G}_V^{R} (q_3) \right]^\pm 
\nn \\
&\quad\qquad 
- k^\alpha q_1^\beta \,   (T_T)^*_{ \beta \alpha} (q_3,q_2)  \,   \left[\bar{G}_V^{R} (q_1) \right]^\pm    \,  \Phi^\dagger (q_2)  \, \Phi (q_3) 
\nn \\
&\quad\qquad 
+ 2 (k q_2)  (q_1 q_3)    \left[\bar{G}_V^R(q_1) \right]^\pm    \left[\bar{G}_V^R(q_2) \right]^\mp  \left[\bar{G}_V^R(q_3) \right]^\pm   
\bigg\}\,,
\end{align}
and 
\begin{align}
P_T^\pm   (k) &=  -4  \, G_F^2   \, 
 \int  \frac{d^4q_1 d^4q_2 d^4q_3}{(2 \pi)^8}  
\,  \delta^{(4)}  (k - q_3  - q_1 + q_2)
\nn \\
&\quad \times  \bigg\{
\pm \frac{1}{2}  \Phi(q_1) \Phi^\dagger (q_2) \Phi (q_3)   \,  (T_T)_{\mu \nu} (q_2,q_1) \, (T_T)^{\mu \nu} (k, q_3)
\nn \\
&\quad\qquad 
\pm  \Phi (q_1)  \left[\bar{G}_V^R(q_2) \right]^\mp  \left[\bar{G}_V^R(q_3) \right]^\pm \,  q_2^\alpha q_3^\beta \,    (T_T)_{\alpha \beta} (k,q_1) 
\nn  \\
&\quad\qquad \mp  
 \left[\bar{G}_V^L(q_1) \right]^\pm  \Phi (q_2)  \left[\bar{G}_V^R(q_3) \right]^\pm \,  q_1^\alpha q_3^\beta \,    (T_T)_{\alpha \beta} (k,q_2) 
\nn  \\
&\quad\qquad \pm
 \left[\bar{G}_V^L(q_1) \right]^\pm  \left[\bar{G}_V^L(q_2) \right]^\mp    \Phi (q_3) \   q_1^\alpha q_2^\beta \,    (T_T)_{ \beta \alpha} (k,q_3) 
\bigg\}\,.
\end{align}
\end{subequations}

\subsection{Frequency projections: general results for loss  and gain potentials}
\label{sect:gain-loss-majorana}

To obtain the Majorana collision term ${\cal C}_M$, we need the positive-frequency ($k^0>0$) integrals of $\hat \Pi^\pm (k^0,\vec{k})$ defined in Eq.~\eqref{eq:C-CbarM}.
Furthermore, we also integrate over $q^0_{1,2,3}$ using the $\d$-functions present in all Green functions, see \eqref{eq:Green-neutrino} and
\eqref{eq:Green-massiveparticle}.

In the following, we will use abbreviations for the various density matrices,  
 i.e. $f_i\equiv f(\vec{q}_i)$, $\bar f_i\equiv \bar f(\vec{q}_i)$, $\bar f^T_i\equiv \bar f^T(\vec{q}_i)$, $f^T_i\equiv f^T(\vec{q}_i)$, and $\phi_i\equiv\phi(\vec{q}_i)$, $\phi_i^T\equiv\phi^T(\vec{q}_i)$, $\phi_i^\dagger\equiv\phi^\dagger(\vec{q}_i)$, $\phi_i^*\equiv\phi^*(\vec{q}_i)$. 
We will, however, omit subscripts $_k$: $f\equiv f(\vec{k})$, $\phi\equiv\phi(\vec{k})$.
Note that all density matrices and distribution functions appear with argument ``$+ \vec{q}_i$''. 
i.e. $f (\vec{q}_i)$ and not  $f (-\vec{q}_i)$,  something we achieve via variable substitution under the integrals.
Furthermore, $f,\,\bar f$ with subscripts $_{(N)},_{(e)}$ indicate them being nucleon and electron (anti)particle distributions (and thus scalars in flavor space) rather than neutrino distributions.
Finally, we write 
\be
\int \widetilde{dq_i}\equiv\int \frac{d^3\vec{q}_i}{2E_i(2\pi)^3}~, 
\label{eq:tildemom}
\ee
where the energy is $E_i\approx\sqrt{(\vec{q}_i)^2}$ for neutrinos (since their masses would give  $O(\epsilon^3)$ contributions in the collision term), 
and $E_i=\sqrt{(\vec{q}_i)^2+M^2}$ for electrons and nucleons with $M=M_e$ and $M=M_N$, respectively.

Below, we give the expressions for the loss potentials $\Pi^{\kappa +}_{R} (\vec{k})$,  $P_T^+(\vec k)$ corresponding to each 
class of processes in the medium.  From these,  the gain potentials $(\Pi^{\kappa}_{R})^-(\vec{k})$ and $P_T^- (\vec k)$ can be obtained as follows, 
\be
(\Pi^{\kappa}_{R})^-(\vec{k})
=(\Pi^{\kappa}_{R})^+(\vec{k})\big|_{f_i \to1 -f_i,   \  \phi_j  \to - \phi_j }~, 
\qquad 
P_T^-(\vec{k})=  P_T^+(\vec{k})\big|_{f_i \to  1-f_i, \ \phi_j \to  - \phi_j}
\ee
for all $f_i$ (all particle species, including barred ones) and $\phi_j$. 
For each class of processes, we also  give below the recipe to obtain 
the antineutrino  potentials $\Pi^{\kappa \pm}_{L}(\vec{k})$ 
from  the neutrino  potentials  $\Pi^{\kappa \pm}_{R}(\vec{k})$.

\subsubsection*{Neutrino-nucleon scattering processes}

Neutrino-nucleon scattering  $\nu (k) N (q_2) \to \nu(q_3)  N(q_1)$  
induces the following contributions to  
the  loss potentials $\Pi_R^{\kappa +} (\vec k)$ and $P_T^{+} (\vec k)$:
\begin{subequations}
\begin{align}
\Pi_R^{\kappa +} (\vec k) &=
 -\frac{4 G_F^2}{|\vec{k}|} \int\widetilde{dq}_{1}\widetilde{dq}_{2}\widetilde{dq}_{3} 
 \, (2\pi)^4  
 \M_{R}(q_1,q_2,q_3,k)   (1\!-\!f_{\!(N),1})f_{\!(N),2}     \left( 1\!-\!f_3  \right)
\,,\\
P_T^+ (\vec k) &= \frac{8 G_F^2}{|\vec{k}|} \int\widetilde{dq}_{1}\widetilde{dq}_{2}\widetilde{dq}_{3}   \, (2\pi)^4  
  (C_V^2 + C_A^2) 
  \M_T(q_1,q_2,q_3,k) f_{(N),2}(1\!-\!f_{(N),1})    \phi_3
\,, \label{eq:PInuN}
\end{align}
\end{subequations}
with
\begin{align}
 \M_{R,L}(q_1,q_2,q_3,k)&=  \d^{(4)}(k\!-\!q_3\!-\!q_1\!+\!q_2)
\,  4 \,   \bigg(\big(C_V^2\!+C_A^2\big)\!\left((q_1q_3)(kq_2)+(q_1 k)(q_2q_3)\right) \nonumber\\
 &\quad -\big(C_V^2\!-C_A^2\big)M_{\!N}^2(q_3 k) \pm 2C_VC_A\big((q_1q_{3}) (kq_2)-(q_2q_{3}) (kq_1)\big)\bigg)
 \,,\nn\\
 \M_T(q_1,q_2,q_3,k)&=  
 \d^{(4)}(k\!-\!q_3\!-\!q_1\!+\!q_2) \, 
 \abs{\vec{k}}  \abs{\vec{q_3}} 
\,  q_1^\m q_2^\n
\,  (T_T)_{\mu \nu} (k,q_3)~, \!\! \!
\label{eq:MT}
\end{align}
where
$(T_T)_{\mu \nu} (k, k')$ is defined in \eqref{eq:LRTtensors}, 
we suppressed the superscripts $^{(N)}$ on the couplings $C_{V,A}$ and all four-momenta are on-shell,
i.e. $q_i^0=E_i$ and thus $\d^{(4)}(k\!-\!q_3\!-\!q_1\!+\!q_2)=\d(E_k\!-\!E_3\!-\!E_1\!+\!E_2)\d^{(3)}(\vec{k}\!-\!\vec{q}_3\!-\!\vec{q}_1\!+\!\vec{q}_2)$.
The antineutrino potentials are obtained by the relation
\be 
(\Pi^{\kappa}_{L})^\pm(\vec{k})=
(\Pi^{\kappa}_{R})^\pm(\vec{k})\big|_{f_i \to   \bar f_i^T,  \  {\cal M}_R \to {\cal M}_L} ~,
\ee
where  ${\cal M}_R \to {\cal M}_L$ amounts to a change of sign in the axial coupling $C_A$.

\subsubsection*{Charged-current processes}

The loss potential term from charged-current neutrino absorption 
$\nu (k) n (q_2) \to   e^- (q_3)  p (q_1)$   is
\be
\Pi_R^{\kappa +} (\vec k)  =- \frac{4 G_F^2}{\abs{\vec{k}}}  \int\widetilde{dq}_{1}\widetilde{dq}_{2}\widetilde{dq}_{3}   (2\pi)^4
 \M_{R}^{CC}(q_1,q_2,q_3,k)
(1\!-\!f_{\!(p),1})   f_{\!(n),2}   ( 1\!-\!f_{(e),3} )  I_e ~, 
\ee
where the flavor projector $I_e$ is defined in Eq.~\eqref{eq:flprojector} and 
\begin{align}
 \M_{R,L}^{CC} (q_1,q_2,q_3,k)&= 4\, \bigg((1+g_A^2)\big((q_3q_1)(kq_2)+(q_3q_2)(kq_1)\big)
 -M_pM_n(1-g_A^2)(kq_3) \nonumber\\
 &\quad\qquad \pm 2g_A\big((q_3q_1)(kq_2)-(kq_1)(q_3q_2)\big)\bigg)\d^{(4)}(k\!-\!q_3\!-\!q_1\!+\!q_2)  \,.
 \label{eq:MECC}
\end{align}
Neutrino absorption and emission does not induce $P_T^\pm (\vec{k})$. 
The antineutrino potentials are obtained by the relation
\be 
(\Pi^{\kappa}_{L})^\pm(\vec{k})=
(\Pi^{\kappa}_{R})^\pm(\vec{k})\big|_{f_n \leftrightarrow f_p,   \ f_{(e)} \to \bar{f}_{(e)},    \ {\cal M}_R^{CC}  \to {\cal M}_L^{CC} }~,
\ee
where again ${\cal M}_R^{CC} \to {\cal M}_L^{CC}$ amounts to a change of sign in the axial coupling $g_A$. 

\subsubsection*{Neutrino-electron processes}

Neutrino electron processes contribute to the loss potentials as follows,
\begin{subequations}
\begin{align}
\Pi_R^{\kappa +} (\vec k) &=
 -\frac{32 G_F^2}{|\vec{k}|} \int\widetilde{dq}_{1}\widetilde{dq}_{2}\widetilde{dq}_{3} 
 \, (2\pi)^4      
\label{eq:PInuea}
  \\*
 &    \times\sum\limits_{I=L,R}\Bigg[
  (1\!-\!f_{(e),1}) f_{(e),2}
Y_{I}(1\!-\!f_3)\bigg(2Y_{I}\M^{R}_{I}(q_1,q_2,q_3,k)-Y_{J\neq I}\M_m(q_1,q_2,q_3,k)\bigg)   
\nonumber\\*
 &\quad + \bar f_{(e),1}(1\!-\!\bar f_{(e),2})  
  Y_{I}(1\!-\!f_3)\bigg(2Y_{I}\M^{R}_{I}(-q_1,-q_2,q_3,k)-Y_{J\neq I}\M_m(-q_1,-q_2,q_3,k)\bigg) 
 \nonumber\\*
 &\quad +(1\!-\! f_{(e),1})(1\!-\!\bar f_{(e),2}) 
 Y_{I}\bar f_3\bigg(2Y_{I}\M^{R}_{I}(q_1,-q_2,-q_3,k)-Y_{J\neq I}\M_m(q_1,-q_2,-q_3,k)\bigg) 
 \Bigg]
\,,\nonumber \\
P_T^+ (\vec k) &= -  \frac{32   G_F^2}{|\vec{k}|} \int\widetilde{dq}_{1}\widetilde{dq}_{2}\widetilde{dq}_{3}   \, (2\pi)^4  
\nonumber \\*
& \qquad \quad
  \times\sum\limits_{I=L,R}\Bigg[
  (1\!-\!f_{(e),1}) f_{(e),2}
Y_{I}(- \phi_3) Y_{I} \M_{T}(q_1,q_2,q_3,k)  
\nonumber\\*
 & \qquad \qquad
 + \bar f_{(e),1}(1\!-\!\bar f_{(e),2})  
  Y_{I}(- \phi_3) Y_{I}  \M_{T}(-q_1,-q_2,q_3,k)  
 \nonumber\\*
 &  \qquad \qquad
 +(1\!-\! f_{(e),1})(1\!-\!\bar f_{(e),2}) 
   Y_{I} \,  \phi_3^T \, Y_{I}  \M_{T}(q_1,-q_2,-q_3,k)
 \Bigg]~, 
\label{eq:PInueb}
\end{align}
\end{subequations}
where 
\begin{align}
 \M^{L}_{I}(q_1,q_2,q_3,k)&=   \left(  \d_{I}^{R}(q_3q_1)(kq_2)+\d_{I}^{L}(q_3q_2)(kq_1)  \right) \, \d^{(4)}(k\!-\!q_3\!-\!q_1\!+\!q_2)
 \,, \nn\\
  \M^{R}_{I}(q_1,q_2,q_3,k)&= \left( \d_{I}^{L}(q_3q_1)(kq_2)+\d_{I}^{R}(q_3q_2)(kq_1) \right) \, \d^{(4)}(k\!-\!q_3\!-\!q_1\!+\!q_2)
 \,, \nn\\
 \M_m(q_1,q_2,q_3,k)&= m_e^2  \,   (kq_3)  \,  \d^{(4)}(k\!-\!q_3\!-\!q_1\!+\!q_2)  
\label{eq:MEelectron}
\end{align}
and  $\M_T (q_1,q_2,q_3,k)$ is defined in \eqref{eq:MT}.   
The first term of the sum in Eq.~\eqref{eq:PInuea} 
stems from neutrino scattering off electrons   ($\nu (k)  e^- (q_2)  \to \nu (q_3) e^- (q_1)$), 
the second  from neutrino scattering off  positrons ($\nu (k)  e^+ (q_1)  \to \nu (q_3) e^+ (q_2)$) , 
and the third ones from neutrino-antineutrino annihilation into electron-positron pairs ($\nu (k)  \bar \nu  (q_3)  \to e^+ (q_2) e^- (q_1)$). 
The antineutrino potentials are obtained by the relation
\be 
(\Pi^{\kappa}_{L})^\pm(\vec{k})=
(\Pi^{\kappa}_{R})^\pm(\vec{k})\big|_{f_i \to   \bar f_i^T,  \    \bar f_i \to  f_i^T,  \  Y_R  \leftrightarrow  Y_L} ~, 
\ee
where $Y_L \leftrightarrow Y_R$ is equivalent to  the replacement  ${\cal M}^R_I \to {\cal M}^L_I$.

\subsubsection*{Neutrino-neutrino processes}

Neutrino-neutrino scattering  $\nu \nu \to \nu \nu$ and neutrino-antineutrino scattering 
$\nu  \bar \nu \to \nu \bar  \nu$ contribute to the  neutrino loss potentials as follows,
\begin{align}
 (\Pi^{\kappa}_{R})^+(\vec{k})
 &= -4\frac{G_F^2}{\abs{\vec{k}}}\int\widetilde{dq}_{1}\widetilde{dq}_{2}\widetilde{dq}_{3}(2\pi)^4\bigg(
                               \!\Big((1-f_1)f_2+\tr\left((1-f_1)f_2\right)\!\Big)(1- f_3)\M (q_1,q_2,q_3,k) \nn\\*
 &\qquad\qquad -\Big(2\phi_1\phi_2^\dagger+\tr\big(\phi_1\phi_2^\dagger\big)\Big)(1-f_3)  \M_T(q_3,k,q_1,q_2)  \nn\\*
 &\qquad\qquad -(1-f_1)\Big(2\phi_2\phi_3^\dagger  +\tr\big(\phi_2\phi_3^\dagger\big) \Big) \M_T(q_1,k,q_2,q_3) \nn\\*
 &\qquad\qquad +2\phi_1 \bar f_2^{T} \phi_3^\dagger\M_T(q_2,k,q_1,q_3)  \nn\\*
  &\quad +\left\{\!\!\begin{array}{c}
                q_{2,3}\to-q_{2,3},\quad
                f_{2,3}\to(1\!-\!\bar f_{2,3})
                ,\quad \bar f^T_{2,3}\to(1\!-\!f^T_{2,3})
                ,\quad \phi_{2,3}\to-\phi_{2,3}^T
                ,\quad \phi_{2,3}^\dagger\to-\phi_{2,3}^*
               \end{array}\!\!\right\} \nonumber\\*
  &\quad +\left\{\!\!\begin{array}{c}
                q_{1,2}\to-q_{1,2},\quad
                f_{1,2}\to(1\!-\!\bar f_{1,2})
                ,\quad \bar f^T_{1,2}\to(1\!-\!f^T_{1,2})
                ,\quad \phi_{1,2}\to-\phi_{1,2}^T
                ,\quad \phi_{1,2}^\dagger\to-\phi_{1,2}^*
               \end{array}\!\!\right\},
\label{eq:nu-nu-loss}
\end{align}
with   $\M_T (q_1,q_2,q_3,k)$ defined in \eqref{eq:MT} and 
\begin{align}
 \M (q_1,q_2,q_3,k)&=4(q_1 q_3)(q_2 k)\d^{(4)}(k\!-\!q_3\!-\!q_1\!+\!q_2)~.
\end{align}
In absence of spin-coherence ($\phi_i \to 0$) 
the first  term in \eqref{eq:nu-nu-loss} encodes loss terms due to  $\nu_k \nu_2 \to \nu_1 \nu_3$, 
while  the terms in the last two lines  in \eqref{eq:nu-nu-loss}  encode the effects of $\nu  \bar \nu \to \nu \bar  \nu$ processes. 
All the remaining terms,  involving  $\phi_i$, arise due to the fact that target neutrinos  in the thermal bath 
can be in coherent linear superpositions of the two helicity states  (see Section~\ref{sect:oneflavor} for a discussion of this point).
The corresponding antineutrino potentials are obtained by the relation
\be 
(\Pi^{\kappa}_{L})^\pm(\vec{k})=
(\Pi^{\kappa}_{R})^\pm(\vec{k})\big|_{f_i \to   \bar f_i^T,  \    \bar f_i \to  f_i^T,  \  \phi_j \leftrightarrow \phi_j^\dagger,   \ {\cal M}_T \to  {\cal M}_T^*} ~.
\ee
and the contributions of neutrino-neutrino processes to the  helicity off-diagonal  loss potentials read
\begin{align}
 P_T^+(\vec{k})&=4\frac{G_F^2}{\abs{\vec{k}}}\int\widetilde{dq}_{1}\widetilde{dq}_{2}\widetilde{dq}_{3}(2\pi)^4 \nn\\
 &\quad\times\bigg(\!\big(\phi_1\phi_2^\dagger-\tfrac12\tr\big(\phi_1\phi^\dagger_2\big)\big)\phi_3\M_{TT}(q_1,q_2,q_3,k)
 -\tfrac12\tr\big(\phi^\dagger_1\phi_2\big)\phi_3\widetilde\M_{TT}(q_1,q_2,q_3,k)
  \nn\\*
 &\qquad\quad +\Big((1-f_1)f_2 +\tfrac12\tr\left((1-f_1)f_2\right) \Big)\phi_3  \M_T(q_1,q_2,q_3,k)     \nn\\*
 &\qquad\quad +\phi_1\Big(\bar f^T_2(1-\bar f^T_3)+\tfrac12\tr\left(\bar f^T_2(1-\bar f^T_3)\right)\Big)  \M_T(q_3,q_2,q_1,k)   \nn\\*
 &\qquad\quad -(1-f_1)\phi_2(1-\bar f^T_3)\M_T(q_1,q_3,q_2,k)   \nn\\*
   &\quad +\left\{\!\!\begin{array}{c}
                q_{2,3}\to-q_{2,3},\quad
                f_{2,3}\to(1\!-\!\bar f_{2,3})
                ,\quad \bar f^T_{2,3}\to(1\!-\!f^T_{2,3})
                ,\quad \phi_{2,3}\to-\phi_{2,3}^T
                ,\quad \phi_{2,3}^\dagger\to-\phi_{2,3}^*
               \end{array}\!\!\right\} \nonumber\\*
  &\quad +\left\{\!\!\begin{array}{c}
                q_{1,2}\to-q_{1,2},\quad
                f_{1,2}\to(1\!-\!\bar f_{1,2})
                ,\quad \bar f^T_{1,2}\to(1\!-\!f^T_{1,2})
                ,\quad \phi_{1,2}\to-\phi_{1,2}^T
                ,\quad \phi_{1,2}^\dagger\to-\phi_{1,2}^*
               \end{array}\!\!\right\},
\end{align}
with   $\M_T (q_1,q_2,q_3,k)$ defined in \eqref{eq:MT} and 
\begin{align}
 \M_{TT} (q_1,q_2,q_3,k)&=\frac12\abs{\vec{q_1}}\abs{\vec{q_2}}\abs{\vec{q}_3}
 \abs{\vec{k}} \, 
 (T_T)_{\mu \nu} (k,q_1) \,  (T_T)^{\mu \nu} (q_2,q_3) \, 
\, \d^{(4)}(k\!-\!q_3\!-\!q_1\!+\!q_2)
 \,,\nn\\
 \widetilde\M_{TT}(q_1,q_2,q_3,k)&=\frac12\abs{\vec{q_1}}\abs{\vec{q_2}}\abs{\vec{q}_3}
 \abs{\vec{k}} \, 
\,  (T_T)_{\mu \nu} (k,q_2) \,  (T_T)^{\mu \nu} (q_1,q_3) \, 
 \,  \d^{(4)}(k\!-\!q_3\!-\!q_1\!+\!q_2)
 \,.
\end{align}

With the gain and loss potentials at hand,  the collision terms $C$ and $C_\phi$ are then assembled according to \eqnref{eq:collisionterms}. 
We present the lengthy results for (some of) the assembled collision terms in  Appendix~\ref{app:results}.

\section{Dirac  collision term}
\label{app:dirac}
 
In this section  we discuss  the structure of the collision terms  ${\cal C}$ and $\bar{\cal C}$  (\ref{eq:C-Cbar})
appearing in the QKEs   (\ref{eq:qkec1})   for Dirac neutrinos and antineutrinos. 
We do not repeat all the steps reported in Section~\ref{sect:derivation}, but simply outline   
how to map the Majorana expressions into the ones relevant for Dirac neutrinos.

First, note that  the self-energy  diagrams in  Fig.~\ref{fig:feynman2} 
for Dirac neutrinos are obtained from the ones in Sect. \ref{sect:diagrams}, that refer to the Majorana case, 
with the following simple changes:
\begin{itemize}
\itemsep=0pt
 \item[(i)] in the weak vertices  one should make the replacement $\gamma_\mu (P_L - P_R)  \to  \gamma_\mu P_L $;
 \item[(ii)]  in Eq.~\eqref{eq:nunutrace} the trace should be multiplied by a factor of 2;
 \item[(iii)]   in Eq.~\eqref{eq:selfCC} the second term (with $\gamma_\mu P_R$ in the vertices) 
should be dropped. 
\end{itemize}

 As a consequence of the different structure of the vertices, the projections over various spinor and  Lorentz components 
of $\hat \Pi^\pm$  (see (\ref{eq:def-Pi-matrix})) simplify greatly.
Using Eq.(\ref{eq:projections-general}) one sees that 
in the Dirac case $\Pi_L$ and $P_T$ vanish. 
Moreover, in the expressions of $\Pi_R$ only the terms proportional to $\bar{G}_V^L$ survive. 
The above simplifications simply reflect the sterile nature of R-handed neutrinos and L-handed antineutrinos.

Results for the neutrino  ($\Pi_R^{\kappa \pm}$) and antineutrino  ($\bar \Pi_R^{\kappa \pm}$)   gain and loss potentials 
are obtained integrating over positive and negative frequencies according to  Eq.~(\ref{eq:Pi-Pibar}). 
The positive  frequency integral is fairly similar to the Majorana one. 
The negative frequency integral can be cast in a simpler form by performing the change of variables $k^0 \to - k^0$, 
leading to 
\begin{subequations}
\begin{align}
\Pi_R^{\kappa \pm}  (\vec{k}) &=  \int_0^\infty   dk^0  \  \Pi_R^{\kappa \pm} (k^0,\vec{k}) \, \delta (k^0 - \abs{\vec{k}}) 
\,,
\\
\bar{\Pi}_R^{\kappa \pm}   (\vec{k}) &=  - \int_0^{\infty}   dk^0  \  \Pi_R^{\kappa \mp} (- k^0,- \vec{k}) \, \delta (k^0 -\abs{\vec{k}})
\,. \label{eq:Pi-Pibar_v3}
\end{align}
\end{subequations}

Finally, performing the matrix multiplications to obtain the various components of  ${\cal C}$  and $\bar{\cal C}$ we obtain the 
following form in terms of $n_f \times n_f$ blocks: 
\begin{align}
 {\cal C} &=\begin{pmatrix}
    C_{LL} & C_{LR} \\
    C_{LR}^\dagger & C_{RR}
 \end{pmatrix}
\,, &
 \bar{\cal C} &=\begin{pmatrix}
    \bar C_{RR} & \bar{C}_{RL} \\
   \bar C_{RL}^\dagger & \bar C_{LL}
 \end{pmatrix}\,,
\label{eq:CDirac}
\end{align}
with 
\begin{subequations}
\begin{align}
C_{LL} &= \frac12  \{ \Pi_{R}^{\kappa + } ,   f_{LL}\}       -  \frac{1}{2}  \{   \Pi_{R}^{\kappa-},  1\!-\!f_{LL}   \}     
\,,\\
C_{RR} &= 0
\,,\\
C_{LR} &= \frac{1}{2} \left(  \Pi_R^{\kappa +}  +  \Pi_R^{\kappa -} \right)  \, f_{LR}  
\,,
\end{align}
\end{subequations}
and 
\begin{subequations}
\begin{align}
\bar C_{RR} &= \frac12  \{ \bar \Pi_{R}^{\kappa + } ,   \bar f_{RR}\}       -  \frac{1}{2}  \{    \bar \Pi_{R}^{\kappa-},  1\!-\! \bar f_{RR}   \}     
\,,\\
\bar C_{LL} &= 0
\,,\\
\bar C_{RL} &= \frac{1}{2} \left( \bar \Pi_R^{\kappa +}  +  \bar  \Pi_R^{\kappa -} \right)  \, \bar  f_{LR}  ~.
\end{align}
\end{subequations}

The collision terms  $C_{RR}$ and $\bar C_{LL}$  vanish because R-handed neutrinos and L-handed antineutrinos 
do not interact in the massless limit that we adopt here (mass effects in the collision term are higher order in the $\epsilon$ counting). 
The gain and loss potentials  $\Pi_R^{\kappa \pm}$  
and $\bar \Pi_R^{\kappa \pm}$   
can be expressed in terms of neutrino density matrices and  distribution functions
of the medium particles  (electrons, etc.), as in the Majorana case. 
In fact, the  expressions for the Dirac case can be obtained from the ones in the Majorana case with the following 
mapping, which we have checked with explicit calculations:

\begin{itemize}
\item    The Dirac neutrino potentials $\Pi_R^{\kappa \pm}$
 are obtained from the Majorana ones by replacing 
$f \to f_{LL}$ and $\bar{f} \to \bar{f}_{RR}$ everywhere.  

\item The Dirac  antineutrino  potentials $\bar \Pi_R^{\kappa \pm}$ 
are in one-to-one correspondence to the Majorana potentials $(\Pi_L^{\kappa \pm})^T$. 
Their expressions are simply obtained from the Dirac neutrino potentials 
$\Pi_R^{\kappa \pm}$ with the following simple changes:
(i)  in the $\nu N$, $\nu e$, and $\nu \nu$ processes replace 
$f_{LL} \to \bar f_{RR}$ and $\bar{f}_{RR}  \to  {f}_{LL}$ everywhere, 
 and flip the signs of the axial couplings ($C_A \to - C_A$ in $\nu N$ terms and ${\cal M}_R \to {\cal M}_L$ in $\nu e$ terms). 
(ii) In the CC processes,  make the replacements $f_e \to \bar{f}_e$,  $f_n \leftrightarrow f_p$,  and 
flip the sign of the axial coupling ($g_A \to - g_A$).

\end{itemize}

\section{One-flavor limit and interpretation of off-diagonal entries}
\label{sect:oneflavor}

We  now specialize to the one-flavor limit and illustrate the structure of the collision term for the two spin degrees of freedom 
corresponding to Majorana neutrino and antineutrino.   
We will  discuss explicitly  only  the simplest process, namely neutrino-nucleon scattering. 
We  provide a simple form for the  various components of the gain and loss potentials $\Pi^{\pm}$ 
in terms of scattering amplitudes of the two spin states (neutrino and antineutrino) off nucleons. 
We also provide a heuristic interpretation of the results for $\Pi^{\pm}$  in terms of changes in occupation
numbers and quantum coherence due to scattering processes in the  medium. 
Finally,  in the limiting case of nearly-forward scattering  we are able to recover earlier results
by Stodolsky and collaborators~\cite{Harris:1980zi,Stodolsky:1986dx,Raffelt:1992uj}.
Note that while we work with spin degrees of freedom, the discussion applies  to the case of any internal degree of freedom. 

\subsection{Scattering amplitudes}

In the collision terms calculated in the previous sections we have set the neutrino mass to zero, as terms proportional 
to the neutrino mass in the collision term would be $O(\epsilon^3)$ in our counting. 
Therefore, when computing neutrino-nucleon scattering from the interaction Lagrangian $\cL_{\n N}$ 
of \eqref{eq:interaction-lagrangian}, we need to use the massless Majorana neutrino fields, $\n_L$, 
which can be expressed as follows  (with $\widetilde{dk}$ defined in \eqref{eq:tildemom}) 
\begin{align}
 \n_{L}(x)&=P_{L}\,\n(x)=\int \widetilde{dk}
 \, \left(u(k,-) a(k,-)e^{-ikx}+v(k,+)a^\dagger(k,+)e^{ikx}\right)
 \,,
\end{align}
in terms of  spinors $v(k,\pm)=u(k,\mp)$ and 
creation  / annihilation operators $a^\dagger (k,\mp)$ and $a (k,\mp)$.  
The $\mp$ label refers to helicity:  negative (L-handed)  helicity corresponds to the neutrino ($\nu_-$), 
while positive helicity (R-handed) to the antineutrino ($\nu_+$). 
The spinors satisfy the following relations, 
\begin{align}
 u(k,\pm)\bar u(k,\pm)&
 =\slashed{k} P_{L/R}\,, \nn\\
 u(k,\pm)\bar u(k,\mp)&=\pm  |\vec{k}|  \frac{i}{4} e^{\pm i\vp}(\hat\kappa\wedge\hat x^\pm)_{\m\n}\s^{\m\n} \, P_{R/L}
 \,,
\label{eq:spinor-relations}
\end{align}
in terms of the basis vectors $\hat \kappa (k)$ and $\hat x^\pm (k)$ (see Appendix~\ref{app:kin}).

The gain and loss terms can be expressed in terms of the following neutrino and antineutrino scattering amplitudes 
(and their conjugates):
\begin{subequations}\label{eq:amplitudesdef}
\begin{align}
A_{\mp} (k)  & \equiv  A  \Big(   \nu_\mp (k)   N(p) \to  \nu_{\mp} (k') N(p')  \Big)
\,,\\
\bar{A}_{\mp} (k)  & \equiv  A  \Big(   \nu_{\mp} (k') N(p')  \to  \nu_\mp (k)   N(p)    \Big) =  A_{\mp} (k)^*  ~.
\end{align}
\end{subequations}
The amplitudes  $A_{\mp} (k)$   depend  also on $k',p,p'$, but to avoid notational clutter we do 
not write  this down explicitly.   From  the interaction Lagrangian $\cL_{\n N}$  of \eqref{eq:interaction-lagrangian}, recalling 
$\Gamma_N^\mu = \gamma^\mu (C_V^{(N)}  -  C_A^{(N)} \gamma_5)$,  one finds:
\begin{subequations}
\begin{align}
A_{-} (k)  & = -   \sqrt{2} G_F \, \bar{u} (k',-) \gamma^\mu u(k,-)  \ \bar{u}_N (p')  \Gamma_N^\mu  u_N (p)  
\,,\\
A_{+} (k)  & =   \sqrt{2} G_F \, \bar{u} (k,-) \gamma^\mu u(k',-)  \ \bar{u}_N (p')  \Gamma_N^\mu  u_N (p)  ~.
\end{align}
\end{subequations}
Note that the scattering processes do not flip the neutrino spin (this effect enters to $O(m_\nu/E_\nu)$). 
In other words,  we consider the  case in which collisions do not change the internal quantum number.   
In the Standard Model this applies to both spin (neglecting neutrino mass)  and flavor. 

Taking the average over the initial and sum over final nucleon polarizations,  i.e. 
\be
\langle  A_\alpha^* A_\beta \rangle =  \frac{1}{2} \sum_{N \ {\rm pol}}     A_\alpha^* A_\beta   \qquad \qquad  \forall \ \alpha,\beta~, 
\ee
using   $\bar u(p,-)\g^\m u(q,-)=\bar u(q,+)\g^\m u(p,+)$,  the relations \eqref{eq:spinor-relations}, 
and the trace identities for gamma matrices,  one can show that: 
\begin{subequations}
\begin{align}
\langle  |A_ \mp(k)|^2 \rangle   &\propto  M_{R/L}  (p',p,k',k), \quad 
\\
\langle   A^*_- (k) A_+ (k)   \rangle   &\propto  M_T  (p',p,k',k),
\end{align}
\end{subequations}
where $M_{R,L,T}$ are given in \eqref{eq:MT}
 (note the proportionality holds modulo the 4-momentum conservation $\delta$-function  in \eqref{eq:MT}). 
These  results  imply  that  
$\Pi_{R,L}^\kappa$ and $P_T$  can be expressed in terms of 
$\langle A_-^*  A_-\rangle$, $\langle A_+^*  A_+ \rangle $, and  $\langle A_-^*  A_+ \rangle$, respectively.

\subsection{Gain and loss potentials in terms of \texorpdfstring{$\nu$  and  $\bar{\nu}$}{nu and nu-bar}  scattering amplitudes}

Keeping track of all factors, we find that  the gain and loss potentials  $\hat \Pi^\pm$ (as per \eqref{eq:def-Pi-matrix}) 
in the one flavor limit can be written  in terms of $\langle A_\alpha^* A_\beta \rangle$ as  follows. 
The gain term is given by:
\begin{align}
 \hat \Pi^-( \vec k)   &= - \frac{1}{|\vec{k}|}  \, \int  \widetilde{dk'} \widetilde{dp} \widetilde{dp'}  
 \ (2 \pi)^4  \delta^{(4)} (k + p  - k' - p') \,  \Big(1 - f_N(p)  \Big)  f_N(p')  
\nonumber  \\ 
 & \qquad \qquad  \times  
 \left(\begin{array}{ccc}
\langle |\bar{A}_-(k)|^2 \rangle  \  f (\vec{k}')    &   & 
\langle  \bar A_- (k)   \bar A_+^* (k) \rangle  \phi (\vec{k}') \\
& & \\
\langle  \bar A_-^* (k)   \bar A_+(k) \rangle  \phi^* (\vec{k}')  &   & 
 \langle |\bar{A}_+(k)|^2 \rangle  \  \bar{f} (\vec{k}')
\end{array}
\right)~, 
\label{eq:pm1f}
\end{align}
while the loss term reads: 
\begin{align}
 \hat \Pi^+(\vec k)   &= - \frac{1}{|\vec{k}|}  \, \int  \widetilde{dk'} \widetilde{dp} \widetilde{dp'}  
 \ (2 \pi)^4  \delta^{(4)} (k + p  - k' - p') \,  f_N(p)   \Big(1 - f_N(p')  \Big)  
\nonumber  \\ 
 & \qquad \qquad  \times  
 \left(\begin{array}{ccc}
\langle |{A}_-(k)|^2 \rangle  \ \left( 1-  f (\vec{k}') \right)    &   & 
\langle   A_-^* (k)    A_+ (k) \rangle   \left( - \phi (\vec{k}') \right) \\
& & \\
\langle   A_- (k)    A_+^* (k) \rangle  \left( - \phi^* (\vec{k}') \right)  &   & 
 \langle |{A}_+(k)|^2 \rangle  \ \left( 1 -  \bar{f} (\vec{k}') \right)
\end{array}
\right)~.
\label{eq:pp1f}
\end{align}
Using \eqref{eq:def-Pi-matrix}  one can easily identify   $\Pi_{R/L}^{\kappa \, \pm} (k)$ and $P_T^\pm (k)$ 
as the diagonal and off-diagonal entries in the above equations.  
Moreover, one can check that the positive-frequency integrals of  \eqref{eq:nucleon-scatt-proj} 
in the one-flavor case reduce to the matrix entries in \eqref{eq:pm1f} and \eqref{eq:pp1f}.

The diagonal entries in the above expressions correspond to the familiar  gain and loss terms for neutrino and antineutrinos ($\nu_\mp (k)$), 
that one could have guessed without the field-theoretic derivation: they are proportional to the square moduli of the scattering  
amplitudes of each state ($|A_{\mp} (k)|^2$).  The off-diagonal entries, however, are  proportional to the products  $A_{-}^* (k) A_+ (k) \, \phi (\vec{k}')$  
and thus are related to interference effects that  arise when initial and final states in a scattering process are given 
by  coherent linear combinations of $\nu_+(k)$, $\nu_-(k)$  and  $\nu_+(k')$, $\nu_-(k')$  ($\phi (\vec{k}') \neq 0$). 

While so far we have phrased our discussion in terms of neutrinos and antineutrinos of the same flavor,  the results generalize to a system 
with any internal degree of freedom, such as flavor, denoted by labels $a,b$. 
Assuming that scattering processes do not change the internal degree of freedom,  i.e.  $A (\nu_a  N \to \nu_b N)  \propto \delta_{ab} A_a$, 
one gets the general structures:
\begin{subequations}
\begin{align}
 \Pi_{ab}^-(k)   &= - \frac{1}{|\vec{k}|}  \, \int  \widetilde{dk'} \widetilde{dp} \widetilde{dp'}   
  \, (2 \pi)^4  \delta^{(4)} (k \!+\! p \!-\! k' \!-\! p') \,  \Big(1 - f_N(p)  \Big)  f_N(p') \, \bar{A}_a (k)  f_{ab} (k')  A_b (k)
\label{eq:pim-gen}
\,,\\
   \Pi_{ab}^+(k)   &=  - \frac{1}{|\vec{k}|}  \, \int  \widetilde{dk'} \widetilde{dp} \widetilde{dp'}  
 \, (2 \pi)^4  \delta^{(4)} (k \!+\! p \!-\! k' \!-\! p') \,  f_N(p)   \Big(1 - f_N(p')  \Big)   \,
    \bar{A}_a (k)  (\id - f (k'))_{ab}  A_b (k) ~,
    \label{eq:pip-gen}
\end{align}
\end{subequations}
in agreement with earlier work on collisional terms for particles with internal degrees of freedom, such as 
color, flavor, and/or spin~\cite{Arnold:1998cy,Botermans:1988xd,Dev:2014laa}.

The above results for $\hat \Pi^\pm (k)$ are derived in the field theoretic context with a well defined set of truncations, 
dictated by our power-counting in $\epsilon$'s.  
In addition, heuristic arguments can help explaining the structure of the gain and loss potentials.  
Let us  discuss $\Pi^-_{ab} (k)$, i.e. the ``gain term''.  
 As we already mentioned,  the off-diagonal terms must be related to interference effects in the scattering, 
arising when the thermal bath contains states that are coherent superpositions of  $|\vec{k},a \rangle$ and $|\vec{k},b \rangle$, 
i.e. states with same momentum but different internal quantum number.  So let us consider the evolution of an initial state 
$ | i \rangle  = c_a (k')  |\vec{k}',a\rangle + c_b (k')  |\vec{k}',b\rangle $, where $\vec{k}'$ represents a generic momentum 
other than the momentum $\vec{k}$ of our ``test'' neutrino. 
Modulo normalizations,   the density matrix associated with this (pure) state reads 
$f_{ab} (k') \propto  c_a (k')  c_b^* (k')$.  
Under S-matrix evolution  the state  $|i\rangle$ evolves into 
$ | f \rangle  =   S | i \rangle   \propto  c_a (k')  \bar{A}_a (k)   |\vec{k},a\rangle + c_b (k') \bar{A}_b (k)   |\vec{k},b\rangle  + \dots$, 
where we used  $ \langle \vec{k},a|  S | \vec{k}', a\rangle \propto  \bar{A}_a (k)$  (see \eqref{eq:amplitudesdef})  and 
 the dots represent states with  $\vec p \neq \vec k$  onto which the  final state $S |i \rangle$ can project. 
So as a net result of  evolving the state  $|i\rangle$, a linear superposition of internal states with momentum $\vec{k}$ is generated.
The change in the density matrix for momentum  $\vec k$ reads $\Delta f_{ab} (k) \propto    f_{ab} (k')  \bar{A}_a (k) \bar{A}_b^* (k)$, 
which has the same structure of  \eqref{eq:pim-gen}.
So we see that $\Pi^-_{ab} (k) \propto \Delta f_{ab} (k)$, i.e. the gain potential is related to the change in occupation number ($a=b$) 
or coherence ($a \neq b$) in the momentum state $\vec{k}$ resulting from scattering 
from all bins $\vec{k}'$ into the bin $\vec k$.

\subsection{Coherence damping}

The coherence damping rate has been estimated in Refs.~\cite{Harris:1980zi,Stodolsky:1986dx,Raffelt:1992uj} 
in the special case of nearly forward scattering, namely $k' \sim k$, and our expression can reproduce their result. 
In fact, for $k' \sim k$ one has  $f_N(p)  (1 - f_N(p')) \sim  f_N(p')  (1 - f_N(p))$. 
Using this result in  \eqref{eq:pm1f} and  \eqref{eq:pp1f} one obtains 
$P_T^- (k)  = - P_T^+ (k)$ (the latter relation holds under the weaker  condition $|\vec{k}| \sim |\vec{k}'|$). 
Inspection of the collision terms $C$, $\bar C$, and $C_\phi$ of  Eqs.~\eqref{eq:CMajorana} and \eqref{eq:collisionterms} 
shows that in this limit  $C(k) = \bar C (k) \simeq 0$.
On the other hand, 
using $\phi (k') \sim \phi (k)$,  one finds  $C_\phi (k)$ in \eqref{eq:collisionterms}   
to be proportional to  $|A_-(k)|^2 + |A_+ (k)|^2  - 2  A_-^*(k) A_+ (k)$.  Recalling that for $k'\sim k$  then 
$A_-^* A_+$ becomes real,  one arrives at the result
\begin{align}
C_\phi (\vec{k}) &= - \Gamma_\phi (\vec k)  \,  \phi (\vec{k})
\,,\nn \\
\Gamma_\phi (\vec k)  &=
     - \frac{1}{|\vec{k}|}  \, \int  \widetilde{dk'} \widetilde{dp} \widetilde{dp'}  
 \, (2 \pi)^4  \delta^{(4)} (k \!+\! p \!-\! k' \!-\! p') \,  f_N(p)   \Big(1 - f_N(p')  \Big)  
 \, \frac{1}{2}   \langle |A_-(k) - A_+ (k)|^2 \rangle
 \,,
\end{align}
which agrees qualitatively  with \cite{Harris:1980zi,Stodolsky:1986dx,Raffelt:1992uj,Raffelt:1996wa}: the damping rate for the coherence $\phi (\vec{k})$ 
is proportional to a statistical average of the square of the difference of the scattering amplitudes 
of the two states.   In the case of neutrinos and antineutrinos, since weak interactions are spin-dependent, 
$A_- - A_+ \neq 0$ and we expect damping of spin coherence with a typical weak-interaction time scale. 
On the other hand,  neutrino-nucleon scattering is flavor blind and therefore does not contribute to damping of flavor coherence 
in the case of nearly forward scattering. 

``Flavor blind'' scattering (i.e. $A_+ = A_-$ ) can still cause coherence damping, as long as the collisions involve energy transfer. 
Assuming for simplicity thermal equilibrium for the ``scatterers'' (the nucleons in our example, so that $f_N(p) = 1/(e^{E_p/T} + 1)$), we find~
\footnote{
For the ``flavor diagonal'' collision term $C(\vec{k})$, in the same limit we obtain
\begin{align*}
C  (\vec{k}) & =  - \frac{1}{|\vec{k}|}  \, \int  \widetilde{dk'} \widetilde{dp} \widetilde{dp'}  
 \, (2 \pi)^4  \delta^{(4)} (k \!+\! p \!-\! k' \!-\! p') \  |A_-(k)|^2 \   f_N(p)   \left(1 - f_N(p')  \right) 
\nn \\
& \quad \times\! \Big\{ f (\vec{k}) (1 - f (\vec{k}')) -  e^ {(E_p - E_{p^\prime})/T}  f (\vec{k}') (1 - f (\vec{k})) 
  +   \left( e^{(E_p - E_{p^\prime})/T}  - 1 \right)  \, {\rm Re} \left( 
\phi (\vec{k})   \phi^* (\vec{k}') \right)  
\Big\}~.
 \end{align*}
}  
\begin{align}
&C_\phi (\vec{k}) =  - \frac{1}{|\vec{k}|}  \, \int  \widetilde{dk'} \widetilde{dp} \widetilde{dp'}  
 \, (2 \pi)^4  \delta^{(4)} (k \!+\! p \!-\! k' \!-\! p') \  |A_-(k)|^2 \   f_N(p)   \Big(1 - f_N(p')  \Big) 
 \\ 
& 
 \quad\;\times\!\Bigg\{ \phi (\vec{k}) -  e^ {(E_p - E_{p^\prime})/T}  \phi (\vec{k}')  
 +   \, \frac{1}{2}  \left( e^{(E_p - E_{p^\prime})/T}  - 1 \right)
\bigg[
\phi (\vec{k})   \left(f(\vec{k}')  + \bar{f}  (\vec{k}') \right)  + 
\phi (\vec{k}')   \left(f(\vec{k})  + \bar{f}  (\vec{k}) \right) 
\bigg]
\Bigg\}
 .
 \nn
 \label{eq:damping-blind}
\end{align}
The  vanishing of  $\int d^3k  C_\phi (\vec{k}) = 0$  (in agreement with Ref.~\cite{Raffelt:1992uj}) 
signals that coherence at the level of the ``integrated'' density matrix is not damped for flavor blind 
interactions. On the other hand, the fact that the individual $C_\phi (\vec{k}) \neq 0$ signals 
that  flavor-blind collisions  ``shuffle'' or transfer coherence between  momentum modes.

\section{Isotropic limit and the early universe}
\label{sect:isotropic}

In this section we revert to the full three-flavor analysis and consider the 
limiting case  of our expressions corresponding to isotropic space, 
that  allows us  to further evaluate analytically the expressions derived earlier in this work.  
The isotropic limit is of  considerable physical interest, as it  applies to the description of the early universe 
(see earlier works~\cite{Hannestad:1995rs,Dolgov:1997mb,Mangano:2005cc,Grohs:2015tfy,Grohs:2015eua}). 
In this setup  we may assume that all $f$'s depend only on the absolute values of the momenta 
(not the angles) and additionally all $\phi$ (for Majorana neutrinos) and $f_{LR}$ (for Dirac neutrinos) vanish, 
thus greatly simplifying our collision terms.
In particular, all collision terms relating to spin-coherence vanish, i.e. $C_\phi=0$ in the Majorana case and $C_{LR} = \bar C_{RL} = 0$ in 
the Dirac case.  

In the Majorana QKEs, the non-vanishing $n_f \times n_f$ blocks of ${\cal C}_M$ in  Eq.~\eqref{eq:CMajorana} are given by:
\begin{subequations}
\begin{align}
 C&=\frac12\left(\{\Pi_{R}^{+,\kappa},f\}-\{\Pi_{R}^{-,\kappa},(1\!-\!f)\}\right)
\,,\\
  \bar{C}^T &= \frac12\left(\{(\Pi_{L}^{\kappa})^+, \bar{f}^T\}-\{(\Pi_{L}^{\kappa})^-,(1\!-\! \bar{f}^T)\}\right)
 \,.
\end{align}
\end{subequations}

In the Dirac QKEs, the non-vanishing $n_f \times n_f$ blocks of ${\cal C}$  and $\bar{\cal  C}$ in  Eq.~\eqref{eq:CDirac}  
are obtained from the  Majorana results as follows:  
\be
C_{LL} =C \Big|_{f \to f_{LL}, \bar{f} \to \bar{f}_{RR}}~, 
\qquad \qquad 
\bar{C}_{RR} =  \bar{C} \Big|_{f \to f_{LL}, \bar{f} \to \bar{f}_{RR}}~.
\ee

The key trick~\cite{Dolgov:1997mb,Dolgov:1998sf}
leading to closed expressions for the collision integrals is to
write the momentum conserving $\d$-function in terms of its Fourier representation
\begin{align}
 \d^{(3)}(\vec{k}-\vec{q}_3-\vec{q}_1+\vec{q}_2)=\int\!\frac{d^3\l}{(2\pi)^3}e^{i\vec{\l}(\vec{k}-\vec{q}_3-\vec{q}_1+\vec{q}_2)}
 =\int\!\frac{r^2_\l dr_\l d(\cos\th_\l)d\vp_\l}{(2\pi)^3}e^{i\vec{\l}(\vec{k}-\vec{q}_3-\vec{q}_1+\vec{q}_2)}
 \,.
\label{eq:frep}
\end{align}
With this result, one 
can integrate out all angles ultimately arriving at an expression with  only two integrals left~\cite{Dolgov:1997mb,Dolgov:1998sf}.
The types of integrals appearing for the $x=\cos\th_i$ are
\begin{align}
 \int_{-1}^1dxe^{iAx}&=\frac2A\sin A\,, &
 \int_{-1}^1dx xe^{iAx}&=\frac{-2i}A\left(\cos A-\frac{\sin A}{A}\right)
 \,.
\end{align}
The integrals over the $\vp_i$ are trivial when aligning the $\hat z$-axis with $\vec{\l}$.

Below, we report our results for the collision term in isotropic environments. 
Each contribution to the collision term has a factorized structure,
 in terms of  weak matrix elements and distribution functions of the ``scatterers'', 
multiplied by a matrix structure involving the neutrino and antineutrino density matrices  $f, f_{1,2,3}$ and $\bar{f}, \bar{f}_{1,2,3}$.

\subsubsection*{Neutrino-nucleon scattering processes}

Neglecting contributions from antinucleons (irrelevant in the early universe in the interesting decoupling region, $T < 20$~MeV) 
the scattering processes  $\nu (k) N (q_2)  \leftrightarrow \nu(q_3)  N(q_1)$  lead to
\begin{align}
 C&=-2\frac{G_F^2}{E_k^2} \int\frac{dE_1dE_2dE_3}{(2\pi)^3}\d(E_k\!-\! E_3\!-\! E_1\!+\! E_2)\nn\\*
 &\quad\times\bigg(   (1\!-\!f_{\!(N),1}) f_{\!(N),2} \,   \Big\{1\!-\! f_3,f\Big\}-     f_{\!(N),1}(1\!-\!f_{\!(N),2} ) \Big\{f_3,1\!-\!f \Big\}\bigg)\nn\\*
 &\quad \times \bigg((C_V+C_A)^2\Big(E_1E_2E_3E_kD_1(q_1,q_2,q_3,k)
 +E_2E_kD_2(q_2,k;q_1,q_3) \nn\\*
 &\qquad\hspace*{4cm} +E_1E_3D_2(q_1,q_3;q_2,k)
 +D_3(q_1,q_2,q_3,k)\Big) \nn\\*
 &\quad\quad +(C_V-C_A)^2\Big(E_1E_2E_3E_kD_1(q_1,q_2,q_3,k)
 -E_1E_kD_2(q_1,k;q_2,q_3) \nn\\*
 &\qquad\hspace*{4cm} -E_2E_3D_2(q_2,q_3;q_1,k)
 +D_3(q_1,q_2,q_3,k)\Big)\nn\\*
 &\quad\quad -M_N^2(C_V^2-C_A^2)\Big(E_3E_kD_1(q_1,q_2,q_3,k)- D_2(q_1,q_2;q_3,k)\Big)\bigg),
\label{eq:CnuNEU}
\end{align}
where $q_{1,2}=\sqrt{E_{1,2}^2-M_N^2}$, $q_{3},k=\sqrt{E_{3,k}^2}$, and $D_{1,2,3}$ are expressions previously discussed by Dolgov, Hansen and Semikoz in~\cite{Dolgov:1997mb} (see also~\cite{Dolgov:1998sf}), and we list them explicitly in \appref{sec:DHSintegrals}.
In the above expressions one recognizes the usual loss and gain terms. 
In the one-flavor limit the anti-commutators become trivial and we recover the standard Boltzmann collision term 
for neutrino-nucleon scattering.   
The antineutrino collision term $\bar{C}^T$ can be obtained from $C$ in \eqref{eq:CnuNEU} with the replacements 
$f_i  \leftrightarrow \bar{f}_i^T$ and $C_A \to - C_A$.

\subsubsection*{Neutrino-electron processes}

The neutrino collision term induced by $\nu$-$e^\pm$ processes is given by
\begin{align}
 C&=-\frac{G_F^2}{E_k^2} \int\frac{dE_1dE_2dE_3}{\pi^3}\d(E_k\!-\! E_3\!-\! E_1\!+\! E_2)
\    (1\!-\!f_{(e),1}) f_{(e),2} \ 
 \times\nonumber\\*
 &\quad \times \bigg(     
 \Big(E_1E_2E_3E_k \, D_1(q_1,q_2,q_3,k)
 +E_2E_kD_2(q_2,k;q_1,q_3)
 +E_1E_3D_2(q_1,q_3;q_2,k) \nn\\*
 &\qquad\qquad +D_3(q_1,q_2,q_3,k)\Big)\Big\{\!   Y_{L}(1\!-\!f_3)Y_{L},f\!\Big\} \nonumber\\*
 &\quad\quad +\Big(E_1E_2E_3E_kD_1(q_1,q_2,q_3,k)
 -E_1E_kD_2(q_1,k;q_2,q_3)
 -E_2E_3D_2(q_2,q_3;q_1,k) \nn\\*
 &\qquad\qquad +D_3(q_1,q_2,q_3,k)\!\Big)\Big\{\!     Y_{R}(1\!-\!f_3)Y_{R},f\!\Big\}\nonumber\\*
 &\quad\quad -\frac{m_e^2}2\Big(E_3E_kD_1(q_1,q_2,q_3,k)- D_2(q_1,q_2;q_3,k)\Big)\sum\limits_{I=L,R}\Big\{\!
  Y_{I}(1\!-\!f_3)Y_{J\neq I},f\!\Big\}\bigg) \nonumber\\*
  &\quad +\left\{\begin{array}{c}
                E_{2,3}\to-E_{2,3},\qquad
                f_{2,3}\to(1-\bar f_{2,3})
               \end{array}\right\} \nonumber\\*
  &\quad +\left\{\begin{array}{c}
                E_{1,2}\to-E_{1,2},\qquad
                f_{1,2}\to(1-\bar f_{1,2})
               \end{array}\right\}
 \nn\\* &\quad +\textrm{gain},
 \label{eq:e-gain-term-res1}
\end{align}
where $q_{1,2}=\sqrt{E_{1,2}^2-m_e^2}$, $q_{3},k=\sqrt{E_{3,k}^2}$, 
``gain'' denotes the corresponding 
gain terms for which the overall sign is flipped and all $f\leftrightarrow(1-f)$ (including barred occurrences of $f_i$),
and the polynomial functions $D_{1,2,3}$ are given in \appref{sec:DHSintegrals}.
The explicit loss  terms in  \eqref{eq:e-gain-term-res1} correspond to $\nu (k)  e^- (q_2)  \to \nu (q_3) e^- (q_1)$ 
scattering. 
The additional loss term expressions indicated implicitly in 
next-to-last and second-to-last lines in  \eqref{eq:e-gain-term-res1} represent 
$\nu (k)  e^+ (q_1)  \to \nu (q_3) e^+ (q_2)$ scattering and $\nu (k)  \bar \nu  (q_3)  \to e^+ (q_2) e^- (q_1) $ pair processes, respectively. 
Note that in these terms,   the sign flips of energies 
affect the energy-conserving delta functions as well as the overall sign of some of the terms proportional to  $D_2$.
The antineutrino collision term $\bar{C}^T$ is obtained from \eqref{eq:e-gain-term-res1} by 
the replacements $f_i  \leftrightarrow \bar{f}_i^T$ and $Y_L \leftrightarrow Y_R$. 
If we  neglect the off-diagonal densities ($f_{a \neq b}=0,  \bar{f}_{a \neq b} =0$), 
the anti-commutators become trivial and we reproduce the 
results of Refs.~\cite{Dolgov:1997mb,Hannestad:1995rs}
for the diagonal entries $C_{aa}$ of the collision term.

\subsubsection*{Charged-current processes}

The processes $\nu (k) n (q_2) \leftrightarrow  e^- (q_3)  p (q_1)$ lead to the neutrino collision term
\begin{align}
 C&=- 2 \frac{G_F^2}{E_k^2} \int\frac{dE_1dE_2dE_3}{(2\pi)^3}\d(E_k\!-\! E_3\!-\! E_1\!+\! E_2)\nn\\*
 &\quad \times \Big(      (1\!-\! f_{(p),1})f_{(n),2}(1\!-\! f_{(e),3}) \  \left\{ I_e, f \right\} -     f_{(p),1}(1\!-\! f_{(n),2})f_{(e),3} \    \left\{ I_e, 1\!-\!f \right\}   \Big)\nn\\*
 &\quad \times\bigg((1+g_A)^2\Big(E_1E_2E_3E_kD_1(q_1,q_2,q_3,k)
 +E_2E_kD_2(q_2,k;q_1,q_3) \nn\\*
 &\qquad\hspace*{3cm} +E_1E_3D_2(q_1,q_3;q_2,k)
 +D_3(q_1,q_2,q_3,k)\Big) \nonumber\\*
 &\quad\quad +(1-g_A)^2\Big(E_1E_2E_3E_kD_1(q_1,q_2,q_3,k)
 -E_1E_kD_2(q_1,k;q_2,q_3) \nn\\*
 &\qquad\hspace*{3cm} -E_2E_3D_2(q_2,q_3;q_1,k)
 +D_3(q_1,q_2,q_3,k)\Big)\nonumber\\*
 &\quad\quad +M_pM_n(g_A^2-1)\Big(E_3E_kD_1(q_1,q_2,q_3,k)- D_2(q_1,q_2;q_3,k)\Big)\bigg),
\label{eq:CccEU}
\end{align}
where $q_{1}=\sqrt{E_{1}^2-M_p^2}$, $q_{2}=\sqrt{E_{2}^2-M_n^2}$, $q_{3}=\sqrt{E_{3}^2-M_e^2}$, $k=\sqrt{E_{k}^2}$,  $D_{1,2,3}$ are  given in \appref{sec:DHSintegrals}, 
and the flavor projector $I_e$ is defined in Eq.~\eqref{eq:flprojector}.

The antineutrino collision term $\bar{C}^T$  induced by the processes 
$\bar{\nu} (k) p (q_2) \leftrightarrow  e^+ (q_3)  n (q_1)$ 
can be obtained from $C$ in \eqref{eq:CccEU} with the replacements 
$f  \leftrightarrow \bar{f}^T$,  $f_{e} \to \bar{f}_e$,   $f_n \leftrightarrow f_p$,  and $g_A \to - g_A$. 
Moreover, $\bar{C}^T$  receives a contribution induced by neutron decay, which
can be obtained from $C$ in \eqref{eq:CccEU} with the replacements 
$f  \leftrightarrow \bar{f}^T$,  $f_{e} \to  1 - f_e$,   $f_n \leftrightarrow f_p$,  and $g_A \to - g_A$, 
and $E_3 \to - E_3$.

\subsubsection*{Neutrino-neutrino processes}

Neutrino scattering off neutrinos and antineutrinos induces the collision term 
\begin{align}
 & C= -2\frac{G_F^2}{E_k^2} \int\frac{dE_1dE_2dE_3}{(2\pi)^3}
\,  \d(E_k\!-\! E_3\!-\! E_1\!+\! E_2)  
\nn\\* 
&\;\times\!
\Bigg(
 \bigg( E_1E_2E_3E_kD_1(q_1,q_2,q_3,k)    +E_2E_kD_2(q_2,k;q_1,q_3)    +E_1E_3D_2(q_1,q_3;q_2,k)   +D_3(q_1,q_2,q_3,k)  \bigg) 
\nn \\
&\quad\quad \times \Big\{\Big(\tr\!\left((1\!-\!f_1)f_2\right)+(1\!-\!f_1)f_2\!\Big)(1\!-\!f_3) \, , \, f \Big\}
\nonumber\\*
 &\;\quad  +
 \bigg(  \! E_1E_2E_3E_kD_1(q_1,q_2,q_3,k)    - E_1 E_kD_2(q_1,k;q_2,q_3)   - E_2E_3D_2(q_2,q_3;q_1,k) + D_3(q_1,q_2,q_3,k)  \!  \bigg)
 \nonumber \\
 &\;\qquad \times   \Big\{\Big(\tr\!\left( \bar f_2   (1\!-\! \bar f_1) \right)+ \bar f_2 (1\!-\ \bar f_1)    \Big)(1\!-\!f_3)   
+        \Big(\tr\!\left( (1  \!- \!f_3)   (1\!-\! \bar f_1) \right)+   (1 \!-\! f_3) (1 \!-\! \bar f_1)  \Big)  \bar f_2   \,  , \,      f     \Big\}
\Bigg)  
\nonumber\\*
 &\quad +  \textrm{gain}\,,
\label{eq:nunuEU}
\end{align}
where all $q_i=\sqrt{E_i^2}$ and ``gain'' denotes the corresponding gain   
terms for which the overall sign is flipped and all $f\leftrightarrow(1-f)$ (including barred occurrences of $f_i$).
The  second and third lines in   \eqref{eq:nunuEU} correspond to 
$\nu (k)  \nu (q_2)  \to \nu (q_3) \nu (q_1)$ 
scattering.  The fourth and fifth lines  in  \eqref{eq:nunuEU} represent 
$\nu (k)  \bar \nu (q_2)  \to   \nu (q_3)  \bar \nu (q_1)$.  
The antineutrino collision term $\bar{C}^T$  induced by the processes 
$\bar \nu \bar \nu \to \bar \nu \bar \nu$ and $  \bar \nu \nu   \to  \bar \nu \nu$
can be obtained from $C$ in \eqref{eq:nunuEU} with the replacements 
$f  \leftrightarrow \bar{f}^T$ (including all  occurrences of $f_i$).
If we  neglect the off-diagonal densities ($f_{a \neq b}=0,  \bar{f}_{a \neq b} =0$), 
the anti-commutators become trivial and we reproduce the 
results of Refs.~\cite{Dolgov:1997mb,Hannestad:1995rs} for the diagonal entries $C_{aa}$ of the collision term.~\footnote{
For the $\nu_a \nu_a \to \nu_a \nu_a$ processes we agree with the overall factor of Ref.~\cite{Dolgov:1997mb} 
which is twice as large as in Ref.~\cite{Hannestad:1995rs}.}

\section{Discussion and conclusions}
\label{sect:conclusion}

In this work we have derived the collision terms entering the  quantum kinetic equations  that describe  
the evolution of   Dirac  or  Majorana neutrinos in a thermal bath.   
We include electroweak processes involving neutrino scattering  off other neutrinos, electrons,  and nucleons, 
as well as $\nu \bar \nu \leftrightarrow e^+ e^-$  pair processes. 

Throughout our analysis we have kept track of both flavor and spin neutrino  degrees of freedom. 
We have first provided general, rather formal, expressions for the collision terms,  
valid in principle for any geometry, including anisotropic environments such as supernovae and 
accretion disks in neutron-star mergers.  
Our results generalize earlier work by Sigl and Raffelt~\cite{Sigl:1992fn}, in which spin coherence effects were neglected. 
When including spin degrees of freedom,   the gain and loss potentials $\Pi^\pm$ become $2 n_f \times 2 n_f$ matrices, 
whose diagonal $n_f \times n_f$ blocks  describe the collisions of each  spin state, 
and whose off-diagonal $n_f \times n_f$ blocks describe interference effects  
in the scattering involving coherent superpositions of the two spin states.  
Since for Dirac neutrinos the ``wrong helicity'' states (R-handed neutrinos and L-handed antineutrinos)
do not interact in the massless limit,   in this case the gain and loss potentials greatly simplify, as only the upper $n_f \times n_f$ block survives 
(see Section~\ref{app:dirac}).
Our results are in qualitative agreement with the ones in Ref.~\cite{Dev:2014laa},
where collision terms involving flavor and helicity coherence have been
studied in the context of kinetic equations for leptogenesis.

The main results of this paper are: 
\begin{itemize}

\item     Within the field-theoretic framework, we have derived general expressions for the neutrino collision terms, 
valid in anisotropic environments.
The lengthy results for the gain/loss potentials are given in Sect.~\ref{sect:derivation}, while the collision terms are presented in Appendix~\ref{app:results}.
Compared to previous literature, new terms involving spin coherence appear.  
After  using  the  constraints from the energy-momentum conservation, 
all the terms (including the ``standard'' ones that do not involve spin)
can be expressed as five-dimensional  integrals. 
These are intractable at the moment in codes describing astrophysical objects, 
but will be required  for  a detailed study of neutrino transport in the future, 
especially to assess the impact of the so-called ``halo'' on collective neutrino oscillations in supernovae~\cite{Cherry:2013lr}.

\item  After  presenting general results, we  have focused on  two limiting  cases  of great  physical interest.
First, in Sect.~\ref{sect:oneflavor}  we  have taken the one-flavor limit and illustrated the structure of the 
collision term for the two spin degrees of freedom   (neutrino and antineutrino in the Majorana case). 
Here we have  provided simple  expressions for the diagonal and off-diagonal entries of the 
gain and loss potential.  As expected, the diagonal terms are proportional to the square moduli of the  
amplitudes describing neutrino and antineutrino scattering off the target particles in the medium ($|A_\mp|^2$).
On the other hand, the off-diagonal terms are proportional to the  product $A_-^*A_+$ 
of scattering amplitudes for  neutrino and antineutrino.  
In this section we have also estimated the damping rate for spin coherence due to neutrinos (antineutrinos) scattering off nucleons, 
under the assumption that collisions involve small energy transfer compared to typical energies of the system.
Finally, we have  shown that coherence ``transfer'' among momentum modes is enforced by the collision term even 
in the case of ``flavor-blind'' interactions, as long as the collisions involve energy transfer.

\item Next, in Sect.~\ref{sect:isotropic} we have   considered  the isotropic limit relevant for the description of neutrinos in the early universe. 
In this  case, following Ref.~\cite{Dolgov:1997mb},  we were able to analytically evaluate most of the  collision terms, 
leaving just two-dimensional integrals  for computational implementation. 
These latter expressions generalize earlier results found in Refs.~\cite{Dolgov:1997mb,Hannestad:1995rs}.
In fact, our results encode the same scattering kernels as in Ref.~\cite{Dolgov:1997mb}, but multiplied by the appropriate 
products of density matrices that realize the ``non-Abelian'' Pauli-blocking first described in Ref.~\cite{Raffelt:1992uj}.
Our flavor-diagonal  collision terms  reproduce the results  of Ref.~\cite{Dolgov:1997mb}. 
The resulting collision terms in the isotropic limit 
are amenable for computational implementation in studies of neutrino 
transport in the early universe, and its impact on primordial lepton number asymmetries and  Big Bang nucleosynthesis. 
\end{itemize}

In summary,  this  work  completes the derivation of neutrino QKEs from field theory 
in general anisotropic environments, started in Ref.~\cite{Vlasenko:2013fja}, by including 
the collision terms.  
While the computational implementation of these collision terms will be challenging, 
here we have provided the  needed theoretical background. 
We have also  gained  insight on the structure of the collision term by discussing 
in some detail the one-flavor limit, relevant for neutrino-antineutrino conversion in compact objects, 
and the isotropic limit,  relevant for the physics of the early universe.

\subsection*{Acknowledgements}
We thank T. Bhattacharya,  J. Carlson,  C. Lee, G. Fuller, E. Grohs, L. Johns, M. Paris,  S. Reddy,  S. Tulin and A. Vlasenko for  insightful discussions.
Special thanks go to E. Grohs for carefully reading the manuscript.
We acknowledge support by the LDRD program at Los Alamos National Laboratory.

\appendix

\section{Kinematics of  ultra-relativistic neutrinos}
\label{app:kin}
\subsection{Basis vectors}
For ultra-relativistic neutrinos of momentum $\vec{k}$,   
it is useful to  express all Lorentz tensors in terms of a basis formed by 
two light-like four-vectors  $\hat\kappa^\mu (k) = (\sgn(k^0), \hat{k})$ and $\hat\kappa'^\mu (k) = (\sgn(k^0), -\hat{k})$
($\hat\kappa\cdot \hat\kappa = \hat\kappa' \cdot \hat\kappa' = 0$,  $ \hat\kappa \cdot \hat\kappa' = 2$) and 
two transverse four vectors  $\hat x_{1,2} (k)$ such that $\hat\kappa \cdot\hat x_i = \hat\kappa' \cdot \hat x_i = 0$ 
and $\hat x_i \cdot\hat x_j = - \delta_{ij}$. 
It also useful to define $\hat x^\pm \equiv \hat x_1 \pm i\hat x_2$ so that $\hat x^+\!\cdot\hat x^-=-2$.
Note, that $k^\m\to -k^\m$ means (with our choice of basis) that $\hat\kappa\to-\hat\kappa$ and $\hat x^\pm\to\hat x^\mp$, 
i.e. $(\hat\kappa\wedge\hat x^\pm)\to-(\hat\kappa\wedge\hat x^\mp)$,
where $(\hat\kappa\wedge\hat x^\pm)_{\m\n}\equiv(\hat\kappa_\mu\hat x^\pm_\nu - \hat\kappa_\nu\hat x^\pm_\mu)$.
This can be seen directly in terms of spherical coordinates:
\begin{align}
 \vec{\hat\kappa}&=\begin{pmatrix}
                   \sin\th\cos\vp\\
                   \sin\th\sin\vp\\
                   \cos\th
                  \end{pmatrix}\,, &
\vec{\hat x}_1&=\begin{pmatrix}
                 \cos\th\cos\vp\\
                 \cos\th\sin\vp\\
                 -\sin\th
                \end{pmatrix}\,, &
\vec{\hat x}_2&=\begin{pmatrix}
                 -\sin\vp\\
                 \cos\vp\\
                 0
                \end{pmatrix} \,, \nonumber\\*
\vec{\hat x}^\pm&=\begin{pmatrix}
                 \cos\th\cos\vp\mp i\sin\vp\\
                 \cos\th\sin\vp\pm i\cos\vp\\
                 -\sin\th
                \end{pmatrix}
\,. \label{eq:basis-in-spherical}
\end{align}
Under parity, the angles change as $\th\to\pi-\th$ and $\vp\to\pi+\vp$.
Therefore, $\cos\th\to-\cos\th$, $\sin\th\to+\sin\th$ and $\cos\vp\to-\cos\vp$, $\sin\vp\to-\sin\vp$, leading to $\vec{\hat\kappa}\to-\vec{\hat\kappa}$, $\vec{\hat x}_1\to+\vec{\hat x}_1$, $\vec{\hat x}_2\to-\vec{\hat x}_2$ and $\vec{\hat x}^\pm\to\vec{\hat x}^\mp$.

Any Lorentz vector $V^\mu$  has light-like and  space-like  components defined as 
$V^\kappa \equiv \hat\kappa \cdot V$ and $V^i \equiv\hat x^i \cdot V$, respectively.
For the components of the derivative operator we adopt the notation
$\partial^\kappa \equiv  \hat\kappa  \cdot  \partial$,  $\partial^i \equiv\hat x^i   \cdot \partial$.

\subsection{Vector and tensor components of two-point functions and self-energies}
\label{app:kin2}

The  decomposition of the neutrino Green function  $G^{(\nu)} (k)$  into independent spinor and  Lorentz structures is  discussed in Ref.~\cite{Vlasenko:2013fja}. 
It takes the form
\begin{align}
G^{(\nu)}  &= \left(\begin{array}{cc} \frac{1}{2}i   G_T^{L,\mu\nu}S^L_{\mu\nu} & G_V^L \cdot\sigma \\  G_V^R \cdot\bar{\sigma} &  \frac{1}{2}i G_T^{R,\mu\nu}S^R_{\mu\nu}\end{array}\right)
\nn\\
     &= \left[  \left(G_V^R\right)^\mu  \gamma_\mu   - \frac{i}{4}   \left( G_T^L \right)^{\mu \nu} \sigma_{\mu \nu} \right]  P_L + 
     \left[   \left( G_V^L \right)^\mu  \gamma_\mu  +  \frac{i}{4}   \left( G_T^R \right)^{\mu \nu} \sigma_{\mu \nu} \right]  P_R~, 
\label{eq:Gdecomposition}
\end{align}
where $\s^\m=(1,\vec{\s})$, $\bar \s^\m=(1,-\vec{\s})$ and $\s^i$ are the usual Pauli matrices.
Additionally,
\begin{align}
 S^L_{\m\n}&=-\frac i4\left(\s_\m\bar\s_\n-\s_\n\bar\s_\m\right) \,, &
 S^R_{\m\n}&=\frac i4\left(\bar\s_\m\s_\n-\bar\s_\n\s_\m\right)
 \,, \nn\\
 -\frac12\s_{\m\n}P_L & =\begin{pmatrix}
   S^L_{\m\n} & 0\\
   0 & 0
  \end{pmatrix}
\,, &
 \frac12\s_{\m\n}P_R & =\begin{pmatrix}
   0 & 0\\
   0 & S^R_{\m\n}
  \end{pmatrix}
  \,.
\label{eq:SLSR}
\end{align}
The first line in Eq.~(\ref{eq:Gdecomposition}) corresponds to the notation of Ref.~\cite{Vlasenko:2013fja}, 
while the second line makes explicit use of the four-dimensional Dirac matrices. 
For additional relations between the four and two component representations, see e.g.~\cite{Dreiner:2008tw}.

Similarly, any self-energy diagram carries spinor indices. It can be written as follows, 
\begin{align}
\hat\Pi&= \left(\begin{array}{cc} \Pi_S+\frac{1}{2}i\Pi_T^{L,\mu\nu}S^L_{\mu\nu} & \Pi_L \cdot\sigma \\ \Pi_R \cdot\bar{\sigma} & \Pi_S^\dagger+\frac{1}{2}i\Pi_T^{R,\mu\nu}S^R_{\mu\nu}\end{array}\right)
\nn\\
     &= \left[ \Pi_S  + \Pi_R^\mu  \gamma_\mu   - \frac{i}{4}   \left( \Pi_T^L \right)^{\mu \nu} \sigma_{\mu \nu} \right]  P_L + 
     \left[ \Pi_S^\dagger   + \Pi_L^\mu  \gamma_\mu  +  \frac{i}{4}   \left( \Pi_T^R \right)^{\mu \nu} \sigma_{\mu \nu} \right]  P_R~, 
\end{align}
where the first line corresponds to the notation of Ref.~\cite{Vlasenko:2013fja}, 
while the second line makes explicit use of the four-dimensional Dirac matrices. 
The vector and tensor spinor components of  $\Pi$ can be isolated  with the following projections: 
\begin{align}
\Pi_{L,R}^\alpha  &=  \frac{1}{2}  \, \Tr \left[ \hat\Pi  \, \gamma^\alpha P_{L,R} \right] \,,
\nn\\
\left( \Pi_T^{L} \right)_{\mu \nu}   &=  \frac{i}{2}  \, \Tr \left[ \hat\Pi  \, \sigma_{\mu \nu}  P_{L} \right] \,,
&
\left( \Pi_T^{R} \right)_{\mu \nu}   &=  -  \frac{i}{2}  \, \Tr \left[ \hat\Pi  \, \sigma_{\mu \nu}  P_{R} \right] ~. 
\end{align}

Furthermore, the  Lorentz vector and tensor objects $\Pi_{L,R}^\alpha$, 
$(\Pi_T^{L,R})_{\mu \nu}$ can be decomposed in terms of the basis vectors $\hat \kappa, \hat \kappa', \hat{x}_{1,2}$. 
The quantities  $\Pi_{L,R}^\kappa$  and $P_T$ that appear in the collision term 
correspond to specific  components  $\Pi_{L,R}^\alpha$ and 
$(\Pi_T^{L,R})_{\mu \nu}$, 
\begin{align}
\left( \Pi_T^R \right)_{\mu \nu} &=  e^{-i\vp}  (\hat \kappa ' \wedge\hat x^{-})_{\mu \nu} \  P_T  \ + \ \dots
\,, &
\left( \Pi_T^L \right)_{\mu \nu} &=  e^{i\vp}  (\hat \kappa ' \wedge\hat x^+)_{\mu \nu} \  P_T^\dagger  \ + \ \dots
\,, \nn\\
\Pi_{L,R}^\mu  &=  \frac{1}{2}  \Pi_{L,R}^\kappa  \,  \hat \kappa^{' \mu}  \ + \ \dots
\end{align}
The components relevant for the collision term are obtained by the contractions
\begin{align}
\Pi_{L,R}^\kappa &= \hat \kappa_\mu \ \Pi_{L,R}^\mu  \,, &
\hat \kappa_\mu &\simeq    \frac{k_\mu}{\abs{\vec{k}}}  
\,, \nn\\
P_T &=  - \frac{e^{i\vp}}{8} \, (\hat \kappa \wedge\hat x^+)^{\mu \nu}  \, \left( \Pi_T^R \right)_{\mu \nu}   \,, &
P_T^\dagger  &=  - \frac{e^{-i\vp}}{8} \, (\hat \kappa \wedge\hat x^-)^{\mu \nu}  \, \left( \Pi_T^L \right)_{\mu \nu}
\,.
\end{align}

So in summary,  to obtain the quantities relevant for the collision term we need 
the following projections (traces are only on spinor indices; $\Pi_{L,R}^\kappa$ and $P_T$ 
are matrices in flavor space): 
\begin{subequations}\label{eq:projections-general-2}
\begin{align}
\Pi_{L,R}^\kappa &= \frac{1}{2}  \hat{\kappa}_\mu  \, \Tr  \left[ \hat\Pi \, \gamma^\mu P_{L,R} \right]
\,,\\
P_T  &=  \frac{ie^{i\vp}}{16} \, (\hat \kappa \wedge\hat x^+)^{\mu \nu}  \,  \, \Tr  \left[ \hat\Pi  \, \sigma_{\mu \nu}  P_{R} \right] 
\,,\\
P_T^\dagger  &=  - \frac{ie^{-i\vp}}{16} \, (\hat \kappa \wedge\hat x^-)^{\mu \nu}  \,  \, \Tr  \left[ \hat\Pi  \, \sigma_{\mu \nu}  P_{L} \right] ~.
\end{align}
\end{subequations}

\subsection{Tensor components and duality properties}
\label{app:kin3}

The projections and traces are greatly simplified by noting the following identities. 
Denote by $T_\pm^{\alpha \beta}$ any  self-dual ($+$)  or anti self-dual  ($-$) antisymmetric tensor, 
with dual tensor   $T^\star$  defined  by 
\be
T^{\star}_{\mu \nu} = \frac{i}{2} \epsilon_{\mu \nu \alpha  \beta} T^{\alpha \beta}~.   
\ee
Then one has:
\begin{align}
\gamma^\mu  \sigma_{\alpha \beta} T_\pm^{\alpha \beta} &=  4 i \,  T_\pm^{\mu \mu'} \gamma_{\mu'} \, P_{R/L}
\,, &
\sigma_{\alpha \beta} T_\pm^{\alpha \beta}  \gamma^\mu   &=  -  4 i \,  T_\pm^{\mu \mu'} \gamma_{\mu'} \, P_{L/R}     ~.
\end{align}
Using the above identities one can perform all the needed projections in a straightforward way 
(at most four gamma matrices and a $\gamma_5$ appear in the traces).

Furthermore, notice also that $S_{L/R}^{\m\n}$ (introduced in \eqref{eq:SLSR})  are (anti)self-dual, i.e.
\begin{align}
 S_L^{\m\n}&=-(S_L^{\m\n})^\star=-\frac{i}{2}\e^{\m\n\r\s}S_{L,\r\s} \,, &
 S_R^{\m\n}&=(S_R^{\m\n})^\star=\frac{i}{2}\e^{\m\n\r\s}S_{R,\r\s}
 \,,
\end{align}
which may be checked by explicit computation using the Lie algebra of the Pauli matrices.
Similarly, the wedge products satisfy the following duality relations
\begin{align}
 (\hat\kappa\wedge \hat x^\pm)^{\m\n}&=\pm \left((\hat\kappa\wedge \hat x^\pm)^{\m\n}\right)^\star=\pm \frac i2\e^{\m\n\r\s}(\hat\kappa\wedge \hat x^\pm)_{\r\s}\,,\nonumber\\
 (\hat\kappa'\wedge \hat x^\pm)^{\m\n}&=\mp  \left((\hat\kappa'\wedge \hat x^\pm)^{\m\n}\right)^\star=\mp \frac i2\e^{\m\n\r\s}(\hat\kappa'\wedge \hat x^\pm)_{\r\s}
 \,,
\end{align}
and from these relations it directly follows that
\begin{subequations}
\begin{align}
 \e_{\a\b\g\m}(\hat\kappa\wedge \hat x^\pm)^{\m\n}&=\pm i\!\left(\d_\a^\n\d_\b^\r\d_\g^\s+\d_\g^\n\d_\a^\r\d_\b^\s+\d_\b^\n\d_\g^\r\d_\a^\s\right)\!(\hat\kappa\wedge \hat x^\pm)_{\r\s}
 \,,\label{eq:wedge-rel-eps} \\
 (\hat\kappa_i\wedge \hat x^-_i)^{\m\n}(\hat\kappa_j\wedge \hat x^+_j)_{\m\n}&=0
 \,, \label{eq:wedge-pm-zero}
\end{align}
\end{subequations}
where we used the notation $\hat{\kappa}_i  \equiv \hat{\kappa} (q_i^\mu)$, 
 $\hat{x}_j^\pm  \equiv \hat{x}^\pm (q_j^\mu)$,  etc.

\section{Green's functions to \texorpdfstring{$O(\epsilon^0)$}{lowest order}  in power counting}
\label{sec:greenfcts}

In this appendix we summarize the form of the two-point functions for  neutrino and matter fields ($e,n,p$)  
to leading order in our power counting, i.e. to $O(\epsilon^0)$. We use these expression  to 
evaluate the collision potentials $\Pi^\pm$ to  $O(\epsilon^2)$.

The collision term involves the $^\pm$ components of the Green's functions, defined in terms of the statistical ($F$) and spectral ($\rho$) functions 
as: 
\be
G^\pm =  - \frac{i}{2}  \, \rho   \ \pm F~.
\ee
Using the fact that to leading order the neutrino spectral function has only vector components, 
\begin{align}
\rho^{(\nu)}_{ab} (k,x) =   2 i \pi  \delta(k^2)  \, {\rm sgn} (k^0)  \, \slashed{k} \, \delta_{ab}~,
\end{align}
we can write (see \eqref{eq:Gdecomposition})
\begin{align}
\left(G_V^{L,R} \right)_{\mu}^\pm (k) &= k_\mu \,    \left(\bar{G}_V^{L,R} \right)^{\pm} (k) 
\,,&
\left( G_T^{L,R} \right)_{\mu \nu} ^\pm &=\pm  \left(F_T^{L,R}\right)_{\mu \nu}
\,,\nn\\
\left(F_T^L \right)_{\mu \nu} &= e^{-i\vp} (\hat{\kappa} \wedge \hat{x}^{-})_{\mu \nu}   \, \Phi~ 
\,,&
\left(F_T^R \right)_{\mu \nu} &= e^{i\vp} (\hat{\kappa} \wedge \hat{x}^{+})_{\mu \nu} \,  \Phi^\dagger ~. 
\end{align}
The explicit form of  $\left(\bar{G}_V^{L,R} \right)^{\pm} (k)$ and $\Phi (k)$ for Majorana neutrinos 
is~\footnote{The Dirac case can be recovered by  the replacements   $f  \to f_{LL}$,   $ \bar{f} \to f_{RR}$ in $\bar{G}_V^L$,  
$\bar{f}^T  \to f_{RR}$,  $ {f}^T \to \bar{f}_{LL}$ in $\bar{G}_V^R$,  
$\phi \to f_{LR}$,  $\phi^T \to \bar{f}^{LR}$ in $\Phi$, and 
$\phi^\dagger \to f_{RL}$,  $\phi^* \to \bar{f}^{RL}$ in $\Phi^\dagger$.}
\begin{subequations}\label{eq:Green-neutrino}
\begin{align}
\left(\bar{G}_V^{L} \right)^+  (k)  &=  2 \pi  \delta (k^2)  \left[   \theta (k^0)    (1 - f (\vec{k}))  -  \theta (-k^0)   \bar{f} (-\vec{k})  \right]  
\,,\\
\left(\bar{G}_V^{L} \right)^-  (k)  &=  2 \pi  \delta (k^2)  \left[   \theta (k^0)   f (\vec{k})   -  \theta (-k^0)  (1 -  \bar{f} (-\vec{k})  )  \right] 
\,,\\
\left(\bar{G}_V^{R} \right)^+  (k)  &=  2 \pi  \delta (k^2)  \left[   \theta (k^0)    (1 - \bar{f}^T (\vec{k}))  -  \theta (-k^0)   f^T  (-\vec{k})  \right]  
\,,\\
\left(\bar{G}_V^{R} \right)^-  (k)  &=  2 \pi  \delta (k^2)  \left[   \theta (k^0)   \bar{f}^T (\vec{k})   -  \theta (-k^0)  (1 - {f}^T (-\vec{k})  )  \right] 
\,,\\
\Phi  (k)  &=  -  2 \pi  \abs{\vec{k}} \delta (k^2)  \left[   \theta (k^0)    \phi (\vec{k})   +   \theta (-k^0)   \phi^T  (-\vec{k})  \right]  ~. 
\end{align}
\end{subequations}
In the above equations the transposition operation acts on flavor indices. 

In summary, the neutrino Green's functions to $O(\epsilon^0)$ can be written as:
\begin{align}
\left( G^{(\nu)} \right)^\pm &=
 \left[  \left(\bar{G}_V^R\right)^\pm  \   k^\mu  \gamma_\mu   \mp   \frac{i}{4}  \,   \Phi   \, e^{-i\vp} (\hat{\kappa} \wedge \hat{x}^{-})^{\mu \nu}  \sigma_{\mu \nu} \right]  P_L
\nn\\
&\quad +
  \left[  \left( \bar{G}_V^L\right)^\pm  \   k^\mu  \gamma_\mu   \pm   \frac{i}{4}  \,   \Phi^\dagger   \, e^{i\vp} (\hat{\kappa} \wedge \hat{x}^{+})^{\mu \nu}  \sigma_{\mu \nu} \right]  P_R
  \,.
\end{align}

Finally, for unpolarized target particles of mass $m$, and spin 1/2 (denoted generically by $\psi$, where $\psi = e,n,p$)  
the   Wigner transformed two-point functions read~\cite{Vlasenko:2013fja,Keldysh:1964ud}:
\begin{align}
G^{(\psi) +} (p) &= 2\pi  \delta (p^2 - m_\psi^2)  (\slashed{p} + m_\psi)   \Big[ \theta(p^0) ( 1 - f_\psi (\vec{p})) - \theta(-p^0) \bar{f}_\psi (-\vec{p}) \Big]
\,,\nn\\
G^{(\psi)-} (p) &= 2\pi  \delta (p^2 - m_\psi^2)  (\slashed{p} + m_\psi)   \Big[ \theta(p^0)   f_\psi (\vec{p})  - \theta(-p^0) (1 -  \bar{f}_\psi (-\vec{p})  ) \Big]
\,,\label{eq:Green-massiveparticle}
\end{align}
where $f_\psi(\vec{p})$ and  $\bar{f}_\psi(\vec{p})$ are the ``target''  particle and antiparticle distributions.
In order to isolate the spinor structure, we  use the notation  $G^{(\psi) \pm} (p)   = (\slashed{p} + m_\psi)  \, \bar{G}^{(\psi)\pm}(p)$.

\section{Collision term for
\texorpdfstring{$\nu$-$N$}{nu-N}, 
 \texorpdfstring{$\nu$-$e$}{nu-e}, and 
charged current processes}
\label{app:results}

In this appendix we present the results for the ``assembled'' collision terms  
 $C$ and $C_\phi$  induced by neutrino-nucleon, 
neutrino-electron, and charged-current processes, 
assembling the gain and loss potentials of Section \ref{sect:gain-loss-majorana}
according to  Eq.~\eqref{eq:collisionterms}. 
We refrain from displaying the expressions for the collision terms induced 
by neutrino-neutrino processes: these are quite lengthy but   
can  be obtained straightforwardly in the same way as for the  other  processes.

\subsubsection*{Neutrino-nucleon scattering processes}
Neutrino-nucleon scattering  $\nu (k) N (q_2) \to \nu(q_3)  N(q_1)$  
induces the following contributions to  
$C$ and $C_\phi$ in \eqref{eq:collisionterms}:
\begin{align}
 C &= -\frac{2G_F^2}{|\vec{k}|} \int\widetilde{dq}_{1}\widetilde{dq}_{2}\widetilde{dq}_{3} 
 \, (2\pi)^4  
 \nn \\
&\quad \times   \Bigg( 
 \M_{R}(q_1,q_2,q_3,k)   \bigg(
 (1\!-\!f_{\!(N),1})f_{\!(N),2}     \left\{ 1\!-\!f_3 ,f\right\}     
  -    f_{\!(N),1}(1\!-\!f_{\!(N),2}  )      \left\{f_3,    1\!-\!f  \right\}     \bigg)
\nn\\
 &\qquad - 4 (C_V^2 + C_A^2) 
\left( f_{(N),2} \!  - \!  f_{(N),1} \right) \,  
\Big(  \M_T(q_1,q_2,q_3,k)    \, \phi_3\phi^\dagger  
 + \M_T^*(q_1,q_2,q_3,k) \, \phi \phi_3^\dagger  
 \Big)     \Bigg)  , 
\label{eq:CnuN}
\\
C_\phi&=
-\frac{2G_F^2}{|\vec{k}|} \int\widetilde{dq}_{1}\widetilde{dq}_{2}\widetilde{dq}_{3}
\, (2\pi)^4
\nn \\ 
&\quad \times  \Bigg(\M_{R}(q_1,q_2,q_3,k  \, )\Big( (1\!-\!f_{\!(N),1})f_{\!(N),2}+ (f_{\!(N),1}\!-\!f_{\!(N),2})f_3   \Big) \, \phi
\nn\\
 &\qquad + \M_{L}(q_1,q_2,q_3,k) \,   \phi \, \Big( (1\!-\!f_{\!(N),1})f_{\!(N),2}+ (f_{\!(N),1}\!-\!f_{\!(N),2})\bar f_3^T\Big)\nn\\
 &\qquad 
 - 4 (C_V^2 + C_A^2)  \M_T(q_1,q_2,q_3,k)\, \Big(    
 (f_{(N),2}\!-\!f_{(N),1})  (f \,   \phi_3+ \phi_3 \, \bar f^T )+2f_{(N),1}(1\!-\!f_{(N),2})\phi_3\Big)\!
 \Bigg) ,
\end{align}
where $\M_{R,L,T} (q_1,q_2,q_3,k)$ given in \eqref{eq:MT} 
and $\M^*_T$ denotes the complex conjugate of $\M_T$,  whose only differences are $\hat x^\pm\leftrightarrow\hat x^\mp$ and the sign of the phase.
 
The contribution to $\bar{C}^T$  in \eqref{eq:collisionterms} can be obtained from $C$ in Eq.~\eqref{eq:CnuN} with the following substitutions:
$f_i  \leftrightarrow \bar{f}_i^T$, $\phi_j  \leftrightarrow \phi_j^\dagger$,  ${\cal M}_R  \leftrightarrow  {\cal M}_L$, and ${\cal M}_T  \leftrightarrow {\cal M}_T^*$.

\subsubsection*{Neutrino-electron processes}
In terms of the matrix elements $\M^{L}_{I}(q_1,q_2,q_3,k)$,  $\M^{R}_{I}(q_1,q_2,q_3,k)$,  $\M_m(q_1,q_2,q_3,k)$ defined in \eqref{eq:MEelectron}
and  $\M_T (q_1,q_2,q_3,k)$ defined in \eqref{eq:MT},  
the collision terms $C$ and $C_\phi$ read
\begin{align}
 C&=-\frac{16G_F^2}{\abs{\vec{k}}} \int\widetilde{dq}_{1}\widetilde{dq}_{2}\widetilde{dq}_{3}(2\pi)^4\nn\\
 &\quad \times\sum\limits_{I=L,R}\bigg(
  (1\!-\!f_{(e),1}) f_{(e),2}    
\, \Big\{  Y_{I}(1\!-\!f_3)\big(2Y_{I}\M^{R}_{I}(q_1,q_2,q_3,k)-Y_{J\neq I}\M_m(q_1,q_2,q_3,k)\big),f\!\Big\}\nonumber\\*
 &\quad\qquad -
  f_{(e),1} (1\!-\!f_{(e),2}) \, \Big\{
Y_{I}f_3\big(2Y_{I}\M^{R}_{I}(q_1,q_2,q_3,k)-Y_{J\neq I}\M_m(q_1,q_2,q_3,k)\big),  1\!-\!f  \!\Big\}\nonumber\\*
  &\quad\qquad -\Big( (f_{(e),2}\!-\!f_{(e),1})Y_I\phi_3Y_I\M_T(q_1,q_2,q_3,k)
\Big)\phi^\dagger
-\phi\Big( (f_{(e),2}\!-\!f_{(e),1})Y_I\phi_3Y_I\M_T(q_1,q_2,q_3,k)
\Big)^{\!\dagger}\bigg)
  \nn\\*
  &\quad +\left\{\begin{array}{c}
                q_{2,3}\to-q_{2,3},\qquad
                f_{2,3}\to(1-\bar f_{2,3})
                ,\qquad \phi_3\to-\phi_3^T
               \end{array}\right\} \nonumber\\*
  &\quad +\left\{\begin{array}{c}
                q_{1,2}\to-q_{1,2},\qquad
                f_{1,2}\to(1-\bar f_{1,2})
               \end{array}\right\}
\label{eq:Ce}
\end{align}
and
\begin{align}
 C_\phi &= -\frac{16G_F^2}{\abs{\vec{k}}} \int\widetilde{dq}_{1}\widetilde{dq}_{2}\widetilde{dq}_{3}(2\pi)^4\sum\limits_{I=L,R}\bigg(\big((1\!-\!f_{(e),1}) f_{(e),2}Y_{I}+(f_{(e),1}\!-\!f_{(e),2})Y_{I}f_3\big)\nn\\*
 &\qquad\qquad\times\big(2Y_{I}\M^{R}_{I}(q_1,q_2,q_3,k)-Y_{J\neq I}\M_m(q_1,q_2,q_3,k)\big)\phi\nn\\*
 &\quad +\phi\big((1\!-\!f_{(e),1}) f_{(e),2}Y_{I}+(f_{(e),1}\!-\!f_{(e),2})Y_{I}\bar f_3^T\big)\big(2Y_{I}\M^{L}_{I}(q_1,q_2,q_3,k)-Y_{J\neq I}\M_m(q_1,q_2,q_3,k)\big)\nn\\*
 &\quad - \Big(\!\big(f(f_{(e),2}\!-\!f_{(e),1})+2f_{(e),1}(1\!-\!f_{(e),2})\big)Y_I\phi_3Y_I+(f_{(e),2}\!-\!f_{(e),1})Y_I\phi_3Y_I\bar f^T\Big)\M_T(q_1,q_2,q_3,k)
 \!\bigg)
 \nn \\*
  &\quad +\left\{\begin{array}{c}
                q_{2,3}\to-q_{2,3},\qquad
                f_{2,3}\to(1-\bar f_{2,3})
                ,\qquad \phi_3\to-\phi_3^T
                ,\qquad \bar f^T_3\to(1-f^T_3)
               \end{array}\right\} \nonumber\\*
  &\quad +\left\{\begin{array}{c}
                q_{1,2}\to-q_{1,2},\qquad
                f_{1,2}\to(1-\bar f_{1,2})
               \end{array}\right\}
               .
\end{align}
%
Note that these expressions  display explicitly the  effect of the process $\nu (k)  e^- (q_2)  \to \nu (q_3) e^- (q_1)$, 
while  the impact of neutrino scattering off positrons and pair processes is obtained by simple substitutions, 
as indicated above. 

The antineutrino collision term $\bar{C}^T$  in \eqref{eq:collisionterms} can be obtained from $C$ in Eq.~\eqref{eq:Ce} with the following substitutions:
$f_i  \leftrightarrow \bar{f}_i^T$, $\phi_j  \leftrightarrow \phi_j^\dagger$,  $Y_R  \leftrightarrow Y_L$,  and ${\cal M}_T  \leftrightarrow {\cal M}_T^*$.

\subsubsection*{Charged-current processes}

The contributions to the collision term from charged-current neutrino absorption and emission
$\nu (k) n (q_2) \leftrightarrow  e^- (q_3)  p (q_1)$  are
\begin{align}
 C &=- \frac{2G_F^2}{\abs{\vec{k}}}  \int\widetilde{dq}_{1}\widetilde{dq}_{2}\widetilde{dq}_{3}(2\pi)^4
 \M_{R}^{CC}(q_1,q_2,q_3,k) 
 \nonumber \\*
& \qquad \times 
\bigg(
(1\!-\!f_{\!(p),1})f_{\!(n),2}(1\!-\!f_{(e),3})
 \left\{  I_e  ,f\right\} 
\  - \     f_{\!(p),1}(1\!-\!f_{\!(n),2})f_{(e),3} 
 \left\{ I_e , 1\!-\!f \right\}
 \bigg)
 \label{eq:collCCapp}
\end{align}
and
\begin{align}
 C_\phi&=- \frac{2G_F^2}{\abs{\vec{k}}}\int\widetilde{dq}_{1}\widetilde{dq}_{2}\widetilde{dq}_{3}(2\pi)^4
 \bigg(
 \M_{R}^{CC}(q_1,q_2,q_3,k) \, 
 \Big( 
 (1\!-\!f_{\!(p),1})f_{\!(n),2}+ (f_{\!(p),1}\!-\!f_{\!(n),2})f_{(e),3}\Big)  \  [ I_e\,  \phi ] 
 \nn\\
 &\quad\quad +\M_{L}^{CC}( q_1,q_2,q_3,k) \,   [\phi \,    I_e] \ \Big( (1\!-\!f_{\!(n),1})f_{\!(p),2}+ (f_{\!(n),1}\!-\!f_{\!(p),2})\bar f_{(e),3}
\Big)
 \bigg)
\end{align}
where the flavor projector $I_e$ is defined in Eq.~\eqref{eq:flprojector} and the matrix elements 
$\M_{R,L}^{CC} (q_1,q_2,q_3,k)$ are given in \eqref{eq:MECC}.

The antineutrino collision term $\bar{C}^T$  induced by the processes 
$\bar{\nu} (k) p (q_2) \leftrightarrow  e^+ (q_3)  n (q_1)$ 
can be obtained from $C$  
in \eqref{eq:collCCapp}
with the replacements   
$f  \to \bar{f}^T$,  $f_{e} \to \bar{f}_e$,   $f_n \leftrightarrow f_p$,  and  ${\cal M}_R^{CC} \to {\cal M}_L^{CC}$ 
($g_A \to - g_A$).

\section{The DHS integrals}
\label{sec:DHSintegrals}
Using Mathematica we find (for all $q_i>0$)
\begin{align}
 &D_1(q_1,q_2,q_3,q_4)=\frac{4}{\pi}\int\limits_0^\infty\frac{d\l}{\l^2}\sin(\l q_1)\sin(\l q_2)\sin(\l q_3)\sin(\l q_4) \nonumber\\*
 &=\frac14\Big(|q_1+q_2+q_3-q_4|+|q_1+q_2-q_3+q_4|+|q_1-q_2+q_3+q_4|+|q_2+q_3+q_4-q_1| \nonumber\\*
 &\qquad -|q_1+q_2-q_3-q_4|-|q_1-q_2+q_3-q_4|-|q_1-q_2-q_3+q_4|-(q_1+q_2+q_3+q_4)\Big),
\end{align}
\begin{align}
  &D_2(q_1,q_2;q_3,q_4)=\frac{4q_3q_4}{\pi}\int\limits_0^\infty\frac{d\l}{\l^2}\sin(\l q_1)\sin(\l q_2)\left[\cos(\l q_3)-\frac{\sin(\l q_3)}{\l q_3}\right]\left[\cos(\l q_4)-\frac{\sin(\l q_4)}{\l q_4}\right] \nonumber\\*
 &= \frac{1}{24} \Big(\left| q_1+q_2-q_3-q_4\right|^3+\left|
   q_1-q_2+q_3-q_4\right|^3-\left| q_1+q_2+q_3-q_4\right|
  ^3+\left| q_1-q_2-q_3+q_4\right|^3 \nonumber\\*
  &\quad\qquad -\left|
   q_1+q_2-q_3+q_4\right|^3-\left| q_1-q_2+q_3+q_4\right|
  ^3-\left| -q_1+q_2+q_3+q_4\right|^3+\left(
   q_1+q_2+q_3+q_4\right)^3\Big) \nonumber\\*
  &\quad +\frac{q_3 q_4}4 \Big(\left| q_1+q_2-q_3-q_4\right|
   -\left| q_1-q_2+q_3-q_4\right| +\left| q_1+q_2+q_3-q_4\right|
   -\left| q_1-q_2-q_3+q_4\right| \nonumber\\*
  &\qquad\qquad +\left| q_1+q_2-q_3+q_4\right|
   -\left| q_1-q_2+q_3+q_4\right| -\left| -q_1+q_2+q_3+q_4\right|
   + q_1+q_2+q_3+q_4 \Big) \nonumber\\*
  &\quad +\frac{q_3\!-\!q_4}8\!
   \left(\sgn\!\left(q_1+q_2+q_3-q_4\right)
   \left(q_1+q_2+q_3-q_4\right)^2+\sgn\!\left(q_1-q_2-q_3+q_4\right)
   \left(q_1-q_2-q_3+q_4\right)^2\right. \nonumber\\*
  &\qquad\qquad \left. -\sgn\!\left(q_1+q_2-q_3+q_4\right)
   \left(q_1+q_2-q_3+q_4\right)^2-\sgn\!\left(q_1-q_2+q_3-q_4\right)
   \left(q_1-q_2+q_3-q_4\right)^2\right) \nonumber\\*
  &\quad +\frac{q_3\!+\!q_4}8\!\left( \sgn\!\left(q_1+q_2-q_3-q_4\right)
   \left(q_1+q_2-q_3-q_4\right)^2+\sgn\!\left(q_1-q_2+q_3+q_4\right)
   \left(q_1-q_2+q_3+q_4\right)^2\right. \nonumber\\*
  &\qquad\qquad \left. -\sgn\!\left(q_1-q_2-q_3-q_4\right)
   \left(-q_1+q_2+q_3+q_4\right)^2-
   \left(q_1+q_2+q_3+q_4\right)^2\right),
\end{align}
{\allowdisplaybreaks
\begin{align}
  &D_3(q_1,q_2,q_3,q_4)=\frac{4q_1q_2q_3q_4}{\pi}\int\limits_0^\infty\frac{d\l}{\l^2}\left[\cos(\l q_1)-\frac{\sin(\l q_1)}{\l q_1}\right]\left[\cos(\l q_2)-\frac{\sin(\l q_2)}{\l q_2}\right]\times \nonumber\\*
 &\qquad\hspace*{3cm} \times\left[\cos(\l q_3)-\frac{\sin(\l q_3)}{\l q_3}\right]\left[\cos(\l q_4)-\frac{\sin(\l q_4)}{\l q_4}\right] \nonumber\\
 &=\frac{1}{120} \Big(q_1^5+q_2^5-5 \left(q_2^2+q_3^2+q_4^2\right) q_1^3-5
   \left(q_2^3+q_3^3+q_4^3\right) q_1^2-5 q_2^3
   \left(q_3^2+q_4^2\right)-5 q_2^2 \left(q_3^3+q_4^3\right)\nonumber\\*
   &\quad\qquad +\left(q_3+q_4\right)^3 \left(q_3^2-3 q_4
   q_3+q_4^2\right)\Big)
  \nonumber\\
  &\quad +\frac{1}{480}\Big(  \left| q_1-q_2-q_3-q_4\right|^5- \left| q_1+q_2-q_3-q_4\right|
  ^5- \left| q_1-q_2+q_3-q_4\right|^5+
   \left| q_1+q_2+q_3-q_4\right|^5 \nonumber\\*
   &\quad \quad - \left|
   q_1-q_2-q_3+q_4\right|^5+ \left|
   q_1+q_2-q_3+q_4\right|^5+ \left|
   q_1-q_2+q_3+q_4\right|^5\Big)
  \nonumber\\
  &\quad +\frac{1}{24}\Big(\left(q_3 q_4+q_2
   \left(q_3+q_4\right)-q_1 \left(q_2+q_3+q_4\right)\right) \left|
   q_1-q_2-q_3-q_4\right|^3 -6 q_1 q_2 q_3 q_4 \left|
   q_1-q_2-q_3-q_4\right| \nonumber\\*
   &\quad \quad +\left(q_1
   \left(q_2+q_3-q_4\right)-q_2 q_3+\left(q_2+q_3\right) q_4\right) \left|
   q_1-q_2-q_3+q_4\right|^3 -6 q_1 q_2 q_3 q_4 \left|
   q_1-q_2-q_3+q_4\right|\nonumber\\*
   &\quad \quad +\left(q_1 q_2+q_3 q_2+q_1
   q_3-\left(q_1+q_2+q_3\right) q_4\right) \left| q_1+q_2+q_3-q_4\right|
  ^3 -6 q_1 q_2 q_3 q_4 \left|
   q_1+q_2+q_3-q_4\right| \nonumber\\*
   &\quad \quad +\left(q_2 \left(q_3-q_4\right)+q_3 q_4+q_1 \left(q_2-q_3+q_4\right)\right)
   \left| q_1-q_2+q_3-q_4\right|^3 -6 q_1 q_2 q_3 q_4 \left|
   q_1-q_2+q_3-q_4\right| \nonumber\\*
   &\quad \quad +\left(q_2
   \left(q_4-q_3\right)-q_3 q_4+q_1 \left(q_2-q_3+q_4\right)\right) \left|
   q_1+q_2-q_3+q_4\right|^3  -6 q_1 q_2 q_3 q_4 \left|
   q_1+q_2-q_3+q_4\right|\nonumber\\*
   &\quad \quad +\left(q_3 q_4-q_2 \left(q_3+q_4\right)+q_1
   \left(-q_2+q_3+q_4\right)\right) \left| q_1-q_2+q_3+q_4\right|^3  -6 q_1 q_2 q_3 q_4 \left|
   q_1-q_2+q_3+q_4\right| \nonumber\\*
   &\quad \quad +  \left(q_2 \left(q_3+q_4\right)-q_3 q_4+q_1 \left(-q_2+q_3+q_4\right)\right)
   \left| q_1+q_2-q_3-q_4\right|^3 -6 q_1 q_2 q_3 q_4 \left|
   q_1+q_2-q_3-q_4\right| 
  \Big)
  \nonumber\\
  &\quad +\frac{1}{96}\bigg(
  \sgn\!\left(q_1+q_2-q_3-q_4\right) 
   \left(q_1+q_2-q_3-q_4\right)^2\Big(q_1^3+3
   q_1 \left(q_2^2-6 \left(q_3+q_4\right) q_2+q_3^2+q_4^2+6 q_3 q_4\right) \nonumber\\*
   &\qquad\qquad +q_2^3 +3 \left(q_2-q_3-q_4\right) q_1^2-\left(q_3+q_4\right)^3-3 q_2^2 \left(q_3+q_4\right)+3 q_2 \left(q_3^2+6
   q_4 q_3+q_4^2\right)\Big) \nonumber\\*
   &\quad \quad +\sgn\!\left(q_1+q_2-q_3+q_4\right)
   \left(q_1+q_2-q_3+q_4\right)^2 \Big(q_3 \left(q_1+q_2-q_3+q_4\right)^2-q_1 \left(q_1+q_2-q_3+q_4\right)^2 \nonumber\\*
   &\qquad\qquad -q_2
   \left(q_1+q_2-q_3+q_4\right)^2 -q_4
   \left(q_1+q_2-q_3+q_4\right)^2+12 q_1 q_2 (q_3- q_4)+12 (q_1+ q_2) q_3 q_4\Big) \nonumber\\*
   &\quad \quad +\sgn\!\left(q_1-q_2+q_3+q_4\right)
   \left(q_1-q_2+q_3+q_4\right)^2 \Big(q_2
   \left(q_1-q_2+q_3+q_4\right)^2 -q_1 \left(q_1-q_2+q_3+q_4\right)^2\nonumber\\*
   &\qquad\qquad -q_3 \left(q_1-q_2+q_3+q_4\right)^2-q_4
   \left(q_1-q_2+q_3+q_4\right)^2+12 q_1 q_2 (q_3+ q_4)+12 (q_2-q_1)
   q_3 q_4\Big) \nonumber\\*
   &\quad \quad +\sgn\!\left(q_1+q_2+q_3-q_4\right)
   \left(q_1+q_2+q_3-q_4\right)^2 \Big(3 \left(q_2^2+6 q_3
   q_2+q_3^2\right) q_4-q_1^3-3 \left(q_2+q_3-q_4\right) q_1^2 \nonumber\\*
   &\qquad\qquad -3
   \left(q_2^2+6 q_3 q_2+q_3^2+q_4^2-6 \left(q_2+q_3\right) q_4\right)
   q_1-\left(q_2+q_3\right)^3+q_4^3-3 \left(q_2+q_3\right) q_4^2\Big) \nonumber\\*
   &\quad \quad +\sgn\!\left(q_1-q_2-q_3+q_4\right)
   \left(q_1-q_2-q_3+q_4\right)^2 \Big(q_1^3+3 \left(q_2^2+6 q_3
   q_2+q_3^2\right) q_4-3 \left(q_2+q_3-q_4\right) q_1^2 \nonumber\\*
   &\qquad\qquad +3
   \left(q_2^2+6 q_3 q_2+q_3^2+q_4^2-6 \left(q_2+q_3\right) q_4\right)
   q_1-\left(q_2+q_3\right)^3+q_4^3-3 \left(q_2+q_3\right) q_4^2\Big) \nonumber\\*
   &\quad \quad +\sgn\!\left(q_1-q_2+q_3-q_4\right)
   \left(q_1-q_2+q_3-q_4\right)^2 \Big(q_1^3-3 q_2 \left(q_3^2-6
   q_4 q_3+q_4^2\right)-3 \left(q_2-q_3+q_4\right) q_1^2 \nonumber\\*
   &\qquad\qquad +3
   \left(q_2^2+6 \left(q_4-q_3\right) q_2+q_3^2+q_4^2-6 q_3 q_4\right)
   q_1-q_2^3+\left(q_3-q_4\right)^3+3 q_2^2 \left(q_3-q_4\right)\Big) \nonumber\\*
   &\quad \quad +\sgn\!\left(q_1-q_2-q_3-q_4\right)
   \left(-q_1+q_2+q_3+q_4\right)^2 \Big(3 q_2 \left(q_3^2+6
   q_4 q_3+q_4^2\right)-q_1^3+3 \left(q_2+q_3+q_4\right) q_1^2 \nonumber\\*
   &\qquad\qquad -3
   \left(q_2^2+6 \left(q_3+q_4\right) q_2+q_3^2+q_4^2+6 q_3 q_4\right)
   q_1+q_2^3+\left(q_3+q_4\right)^3+3 q_2^2 \left(q_3+q_4\right)\Big)
  \bigg).
\end{align}
As discussed in~\cite{Dolgov:1997mb}, there are four different cases of physical interest for which these expressions simplify considerably.
They are listed in (A.15)-(A.25) of that paper and can be checked explicitly using e.g. Mathematica.
For completeness, let us repeat those expressions here:
}

Assuming for all cases (without loss of generality) that $q_1>q_2$ and $q_3>q_4$, we have~\cite{Dolgov:1997mb}:
\begin{description}
 \item[Case 1: \quad] $q_1+q_2>q_3+q_4\,,\quad q_1+q_4>q_2+q_3$ and $q_1\le q_2+q_3+q_4$
 \begin{subequations}
 \begin{align}
  D_1&=\frac12\left(q_2+q_3+q_4-q_1\right)
  \,,\\
  D_2&=\frac{1}{12}\left((q_1-q_2)^3+2(q_3^3+q_4^3)-3(q_1-q_2)(q_3^2+q_4^2)\right)
  \,,\\
  D_3&=\frac{1}{60}\Big(q_1^5-5q_1^3q_2^2+5q_1^2q_2^3-q_2^5-5q_1^3q_3^2+5q_2^3q_3^2+5q_1^2q_3^3+5q_2^2q_3^3 \nn\\*
  &\quad\qquad -q_3^5-5q_1^3q_4^2+5q_2^3q_4^2+5q_3^3q_4^2+5q_1^2q_4^3+5q_2^2q_4^3+5q_3^2q_4^3-q_4^5\Big)
  \,. \label{eq:case1-D3}
 \end{align}
 \end{subequations}
 The unphysical case $q_1> q_2+q_3+q_4$ yields $D_1=D_2=D_3=0$ here.

 \item[Case 2: \quad] $q_1+q_2>q_3+q_4$ and $q_1+q_4<q_2+q_3$
 \begin{subequations}
 \begin{align}
  D_1&=q_4
  \,,\\
  D_2&=\frac{q_4^3}{3}
  \,,\\
  D_3&=\frac{q_4^3}{30}\left(5q_1^2+5q_2^2+5q_3^2-q_4^2\right)
  \,.
 \end{align}
 \end{subequations}

 \item[Case 3: \quad] $q_1+q_2<q_3+q_4\,,\quad q_1+q_4<q_2+q_3$ and $q_3\le q_1+q_2+q_4$
 \begin{subequations}
 \begin{align}
  D_1&=\frac12\left(q_1+q_2+q_4-q_3\right)
  \,,\\
  D_2&=\frac{1}{12}\left(-(q_1+q_2)^3-2q_3^3+2q_4^3+3(q_1+q_2)(q_3^2+q_4^2)\right)
  \,,
 \end{align}
 \end{subequations}
 and $D_3$ equals \eqnref{eq:case1-D3} with variables $q_1\leftrightarrow q_3$, $q_2\leftrightarrow q_4$ exchanged.
 The unphysical case $q_3> q_1+q_2+q_4$ yields $D_1=D_2=D_3=0$ here.

 \item[Case 4: \quad] $q_1+q_2<q_3+q_4$ and $q_1+q_4>q_2+q_3$
 \begin{subequations}
 \begin{align}
  D_1&=q_2
  \,,\\
  D_2&=\frac{q_2}{6}\left(3q_3^2+3q_4^2-3q_1^2-q_2^2\right)
  \,,\\
  D_3&=\frac{q_2^3}{30}\left(5q_1^2+5q_3^2+5q_4^2-q_2^2\right)
  \,.
 \end{align}
 \end{subequations}

\end{description}

\printbibliography

\end{document}